\newif\ifboyscout                         
\boyscouttrue 
\newif\ifpreparepdf                       
\preparepdftrue 

\documentclass[final,leqno,onefignum,onetabnum]{siamltexmm}

\ifboyscout 
  \newcommand{\PC}[1]{$\footnotemark\footnotetext{Predrag: #1}$}
  
  \newcommand{\Xiong}[2]{$\footnotemark\footnotetext{XD #1: #2}$} 
  
\else 
  \newcommand{\PC}[1]{}
  
  \newcommand{\Xiong}[2]{}{} 
  
\fi  

\ifpreparepdf 

\else  

\fi

\newcommand{\cLv} {covariant vector}  
\newcommand{\CLv} {Covariant vector}  
\newcommand{\cLvs} {covariant vectors}  
\newcommand{\CLvs} {Covariant vectors}  
\newcommand{\entangled} {entangled}

\newcommand{\psd}{periodic Schur decomposition}
\newcommand{\Psd}{Periodic Schur decomposition}
\newcommand{\prsf}{periodic real Schur form}

\newcommand{\pse}{periodic Sylvester equation}
\newcommand{\Pse}{Periodic Sylvester equation}

\newcommand{\pqr}{periodic QR algorithm}
\newcommand{\Pqr}{Periodic QR algorithm}
\newcommand{\ped}{periodic eigendecomposition}
\newcommand{\Ped}{Periodic eigendecomposition}
\newcommand{\ps}[2]{\mathbf{#1}^{(#2)}}
\newcommand{\Rve}[1]{v_{#1}} 
\newcommand{\Jve}[2][0]{\ensuremath{{\bf e}_{#2}^{(#1)}}} 
\newcommand{\Fv}{Floquet vector}

\newcommand{\refeq}  [1] {(\ref{#1})}

\newcommand{\rf}     [1] {~\cite{#1}}
\newcommand{\refref} [1] {ref.~\cite{#1}}

\newcommand{\refsect}[1] {sect.~\ref{#1}}

\newcommand{\refSect}[1] {Sect.~\ref{#1}}

\newcommand{\reffig} [1] {figure~\ref{#1}}

\newcommand{\refFig} [1] {Figure~\ref{#1}}

\newcommand{\reftab} [1] {table~\ref{#1}}
\newcommand{\refTab} [1] {Table~\ref{#1}}

\newcommand{\beq}{\begin{equation}}

\newcommand{\eeq}{\end{equation}}
\newcommand{\ee}[1] {\label{#1} \end{equation}}
\newcommand{\bea}{\begin{eqnarray}}

\newcommand{\eea}{\end{eqnarray}}
\newcommand{\barr}{\begin{array}}
\newcommand{\earr}{\end{array}}

\newcommand{\cGLe}{complex Ginzburg-Landau equation}

\newcommand{\norm}[1]{\left\Arrowvert \, #1 \, \right\Arrowvert}
\newcommand{\KS}{Kuramoto-Siva\-shin\-sky}
\newcommand{\KSe}{Kuramoto-Siva\-shin\-sky equation}
\newcommand{\statesp}{state space}

\newcommand{\stabmat}{stability matrix}     
\newcommand{\JacobianM}{Jacobian matrix} %
\newcommand{\JacobianMs}{Jacobian matrices}  %
\newcommand{\po}{periodic orbit}

\newcommand{\rpo}{rela\-ti\-ve periodic orbit}

\newcommand{\eqva}{equi\-lib\-ria}

\newcommand{\reals}{\mathbb{R}}

\newcommand{\sign}[1]{\sigma_{#1}}
\newcommand{\transp}[1]{{#1}{}^\top}

\newcommand{\ssp}{\ensuremath{x}}                
\newcommand{\vel}{\ensuremath{v}}   
\newcommand\xInit{{\ssp_0}}        
\newcommand{\zeit}{\ensuremath{t}}  
\newcommand{\Mvar}{\ensuremath{A}}  
\newcommand{\jMps}{\ensuremath{J}}   
\newcommand\period[1]{{\ensuremath{T_{#1}}}}         
\newcommand{\cycle}[1]{{\ensuremath{\overline{#1}}}}
\newcommand{\jEigvec}[1][]{\ensuremath{{\bf e}_{#1}}} 
\newcommand{\ExpaEig}{\ensuremath{\Lambda}}
\newcommand{\Lyap}{\ensuremath{\lambda}}            
\newcommand{\eigExp}[1][]{
     \ifthenelse{\equal{#1}{}}{\ensuremath{\lambda}}{\ensuremath{\lambda_{#1}}}}
\newcommand{\eigRe}[1][]{
     \ifthenelse{\equal{#1}{}}{\ensuremath{\mu}}{\ensuremath{\mu_{#1}}}}
\newcommand{\eigIm}[1][]{
     \ifthenelse{\equal{#1}{}}{\ensuremath{\omega}}{\ensuremath{\omega_{#1}}}}

\newcommand{\etc}{{etc.}}       

\usepackage{amsmath,amssymb}
\usepackage{amsfonts}
\usepackage{graphicx}
\usepackage{subfigure}
\usepackage{dsfont}
\usepackage{mathrsfs}

\usepackage{ifthen}
\usepackage{tikz}
\usetikzlibrary{calc}
\usetikzlibrary{decorations.pathreplacing}
\graphicspath{{./figs/}}

\title{
Periodic eigendecomposition and its application to
Kuramoto-Sivashinsky system
}

\author{
  Xiong Ding\footnotemark[2]
  \and
  Predrag Cvitanovi\'{c}\footnotemark[2]
}

\begin{document}
\maketitle
\newcommand{\slugmaster}{%
\slugger{siads}{xxxx}{xx}{x}{x--x}}

\renewcommand{\thefootnote}{\fnsymbol{footnote}}
\footnotetext[2]{
Center for Nonlinear Science,
School of Physics,
Georgia Institute of Technology, Atlanta, GA 30332
(\href{mailto:xding@gatech.edu}{xding@gatech.edu},
\href{mailto:predrag.cvitanovic@physics.gatech.edu}{predrag.cvitanovic@physics.gatech.edu}).
} 

\begin{abstract}
\Ped, to be formulated in this paper, 
is a numerical method to compute Floquet spectrum and \Fv s along
periodic orbits in a dynamical system. It is rooted in 
numerical algorithms advances in computation 
of `\cLvs'\footnote{In literature, it is termed
``covariant Lyapunov vectors'', but we prefer to use \cLvs\ in this paper.} 
of the linearized flow along an ergodic trajectory in a chaotic
system. Recent research on \cLvs\ strongly strongly suggests that the 
physical dimension of
inertial manifold of a dissipative PDE can be
characterized by a {finite number} of `{\entangled} modes', dynamically
isolated from the residual set of transient degrees of
freedom. We anticipate that \Fv s display similar properties as \cLvs.
In this paper we incorporate \psd\ to the computation of dynamical 
\Fv s, compare it with other
methods, and show that the method can yield the full Floquet spectrum of
a periodic orbit at every point along the orbit to high accuracy.
Its power, and in particular its ability to resolve eigenvalues whose
magnitude differs by hundreds of orders magnitude, is demonstrated by
applying the algorithm to computation of the full linear stability
spectrum of several periodic solutions in one dimensional
Kuramoto-Sivashinsky flow.
\end{abstract}


\pagestyle{myheadings}
\thispagestyle{plain}
\markboth{XIONG~DING AND PREDRAG~CVITANOVI\'C}{PERIODIC EIGENDECOMPOSITION}

\section{Introduction}
\label{sect:intro}

In dissipative chaotic dynamical systems, the decomposition of the
tangent space of invariant subsets into 
stable, unstable and center subspaces is important for analyzing the
geometrical structure of the solution field\rf{guckb}. 
For equilibrium points, 
the task is quite simple, which is reduced to the eigen-problem of
a single stability matrix, but the scenario is much more difficult
for complex structures, such as periodic orbits and invariant tours, 
since the expanding/contracting rate in high dimensional systems usually 
span a large order of magnitude. Actually, in literature, two different
algorithms are capable of resolving this problem partially originated 
from different settings. The first candidate is \cLv\ 
algorithm\rf{GiChLiPo12, KuPa12, WoSa07}.
It is designed to stratify the Oseledets subspaces\rf{ruelle79} 
corresponding to 
the hierarchy of Lyapunov exponents along a long non-wandering orbit 
on the attractor. \CLv s attract a lot of attention in the past few 
years. They
turn out to be a useful tool for physicists 
to investigate the dynamical properties of the system, such as
hyperbolicity degree\rf{Bosetti2010a, InKoTaYa12, Kuptsov13} and the geometry of 
inertial manifold\rf{TaGiCh11, YaRa11, YaTaGiChRa08}. 
For our
interest in \po s, it produces 
Floquet spectrum and \Fv s. The second candidate is called 
\psd (PSD)\rf{Bojanczyk92theperiodic}, and
was brought up to compute the eigenvalues of the product of a 
sequence of matrices without forming the product explicitly. This is suitable
for solving the eigenvalue problem in tangent space because the 
fundamental matrix in tangent space can be formed as a product of its shorter-time
pieces. However, in its original form, PSD are only capable of computing 
eigenvalues but not eigenvectors. Also PSD seems not well known to the 
physics community.

In this paper, we unify these two methods for
computing Floquet spectrum and \Fv s along periodic orbits or invariant 
tori, and name it after \emph{\ped}. Special attention is exerted to complex conjugate \Fv s.
There are two stages in the process of this algorithm,
each of which can be accomplished by two different methods, so
we study performance of four different algorithms in all. Also it turns
out that the covariant vectors algorithm reduces to one of them when
applied to \po s.

The paper is organized as follows. \refSect{sect:dynamics} describes
briefly the nonlinear dynamics motivation for undertaking this project,
and reviews two existing algorithms related to our
work. Readers interested only in the algorithms itself can skip this part.
We describe the
computational problem in \refsect{sect:problem}. In  \refsect{sect:psd}
we deal with the first stage of \ped, and then show that both the \pqr\
and simultaneous iteration are capable of achieving \psd.
\refSect{sect:eigenvec} introduces power iteration and reordering as two
practical methods to obtain all eigenvectors. In \refsect{sect:error} we
compare the computational effort required by different methods, and
\refsect{sect:applic} applies \ped\ to \KSe, an example which illustrates
method's effectiveness.

\section{Dynamics background and existing algorithms}
\label{sect:dynamics}

The study of dynamical systems is trying to understand 
the statistical properties of the system and the geometrical structure 
of the global attractor. As we will see, periodic orbits plays an important
role in answering both questions. For dissipative systems, 
orbits typically land onto an invariant subset, called global attractor,
after a transient period, and if the system is chaotic, the attractor is
a strange attractor which contains a dense set of 
periodic orbits. The chaotic deterministic flow on strange attractor can be
visualized as a walk chaperoned by a hierarchy of unstable invariant
solutions (\eqva, \po s) embedded in the attractor. An
ergodic trajectory shadows one such invariant solution for a while, is
expelled along its unstable manifold, settles into the neighborhood of
another invariant solution for a while, and continues in this way forever.
Together, the infinite set of these unstable invariant solutions forms
the skeleton of a chaotic attractor, and in fact spatiotemporal averages,
such as deterministic diffusion coefficients, energy dissipation rate,
Lyapunov exponents, \etc\ can be
accurately calculated as a sum over periodic orbits weighted by products of their
unstable Floquet multipliers\rf{Christiansen97, DasBuch}. 
This is one reason we study
the algorithm of computing Floquet spectrum in this paper.

On the other hand, strange attractor is usually a fractal subset with 
unsmooth surface, and is not handy to analyze. So this motivates the 
formulation of concept of inertial manifold\rf{infdymnon}, 
which contains the 
global attractor but is integer-dimensional and exponentially attractive.
The existence of inertial manifold has been proved for many 
dissipative dynamical systems\rf{infdymnon}, 
but the mathematical proof shed little knowledge 
about the dimension of inertial manifold. Although, upper bounds are persistently
improved for some systems, such as \KSe\rf{jolly_evaluating_2000, Robinson-PLA1994}, 
they are far from being tight and give limited 
hint for suitable mode truncation in numerical simulations. 
Recently, however, there is strong numerical
evidence\rf{TaGiCh11, YaTaGiChRa08} that the long-time chaotic (turbulent)
dynamics of at least two spatially extended systems, \KS\ and complex
Landau-Ginzburg, is confined to an inertial manifold that is everywhere
locally spanned by a \emph{finite number} of `{\entangled}' modes,
dynamically isolated from the residual set of {isolated}, transient
degrees of freedom. {\CLvs} exhibit an approximate orthogonality between
the `{\entangled}' modes and the rest, the `{isolated}' modes. These
results suggest that for a faithful numerical integration of dissipative
PDEs, a finite number of {\entangled} modes should suffice, and that
increasing the dimensionality beyond that merely increases the number of
{isolated} modes, with no effect on the long-time dynamics.
This work has been made possible by advances in algorithms for
computation of large numbers of
`{\cLvs}'\rf{GiChLiPo12, ginelli-2007-99, KuPa12, PoGiYaMa06, TaGiCh09, WoSa07}.
While these studies offer strong evidence for finite dimensionality of
inertial manifolds of dissipative flows, they are based
on numerical simulations of long ergodic trajectories and they yield no
intuition about the geometry of the attractor. That is attained by
studying the hierarchies of unstable \po s, invariant solutions which,
together with their Floquet vectors, provide an effective description  of
both the local hyperbolicity and the global geometry of an attractor
embedded in a high-dimensional \statesp. Motivated by the above studies
of {\cLvs}, we formulate in this paper a {\ped} algorithm suited to
accurate computation of Floquet vectors of unstable \po s.

\subsection{Linear stability}
\label{sect:LinStab}

Now, we turn to the definition of Floquet exponents and Floquet vectors.
Let the flow of a autonomous continuous system be described by
$\dot{\ssp} = \vel(\ssp) $, $\ssp \in \reals^n$
and the corresponding time-forward trajectory
starting from $\xInit$ is
$\ssp(\zeit)=f^{\zeit}(\xInit)$.  In the linear
approximation, the deformation of an infinitesimal neighborhood of
$\ssp(\zeit)$ (dynamics in tangent space) is governed by the
\JacobianM\ (fundamental matrix)
$\delta x(\xInit,\zeit)=\jMps^\zeit(\xInit)\,\delta x(\xInit,0)$,
where $\jMps^{\zeit}(\xInit) = \jMps^{\zeit-\zeit_{0}}(\xInit,\zeit_{0})
= {\partial f^{\zeit}(\xInit)}/{\partial\xInit}$.
\JacobianM\ satisfies the semi-group multiplicative property (chain rule)
along an orbit,
\begin{equation}
\jMps^{\zeit-\zeit_{0}}(x(\zeit_{0}) ,\zeit_{0})
=
\jMps^{\zeit-\zeit_{1}}(x(\zeit_{1}),\zeit_{1})
\jMps^{\zeit_{1}-\zeit_{0}}(x(\zeit_{0}),\zeit_{0})
\,.
\label{eq:xjacobian}
\end{equation}
For a periodic point
$\ssp$ on orbit $p$ of period \period{p},
$\jMps_p=\jMps^{\period{p}}(\ssp)$ is called the Floquet matrix
(monodromy matrix) and its
eigenvalues the Floquet multipliers $\ExpaEig_{j}$.
A {Floquet multiplier} is a dimensionless ratio of the final/initial
perturbation along the $j_{th}$ eigen-direction. It is an intrinsic, local
property of a smooth flow, invariant under all smooth coordinate
transformations. The associated
Floquet vectors $\jEigvec[j](\ssp)$,
$\jMps_p\,\jEigvec[j]=\ExpaEig_{j}\jEigvec[j]$, define the invariant
directions of the tangent space at the periodic point
$\ssp=\ssp(\zeit)\in p$. Evolving small initial perturbation aligned with
a Floquet direction will generate the corresponding unstable manifold along
the \po. Floquet multipliers are either real,
$\ExpaEig_{j} = \sign{j}|\ExpaEig_{j}|$, $\sign{j}\in\{1,-1\}$, or form
complex pairs, $\{\ExpaEig_{j},\ExpaEig_{j+1}\} =
\{|\ExpaEig_{j}|\exp(i\theta_j),|\ExpaEig_{j}|\exp(-i\theta_j)\}$, $0
<\theta_j <\pi$. The real parts of
Floquet exponents $\eigRe[j] = (\ln|\ExpaEig_{j}|)/\period{p}$
describe the mean contraction or
expansion rates per one period of the orbit.
The \JacobianM\ is naively obtained numerically by
integrating the {\stabmat}
\begin{equation}
  \label{eq:tangentDynamics}
  \frac{d {\jMps^{\zeit}}}{d \zeit}
  = \Mvar(\ssp) \jMps^{\zeit}
  \,,\quad\text{with}\quad \Mvar(\ssp)
  = \frac{\partial \vel(\ssp)}{\partial \ssp}
\end{equation}
along the orbit.
However, it is almost certain that this process will overflow or
underflow at exponential rate as the system evolves or the
resulting Jacobian is highly ill-conditioned. Thus,
accurate calculation of expansion rate
is not trivial for nonlinear systems, especially for
those that evolve in a high dimensional space. In such cases, the
expansion/contraction rate can easily range over many orders of magnitude,
which raises a challenge to formulating an effective algorithm to tackle this
problem. However, the semi-group property \refeq{eq:xjacobian} enables
us to factorize the \JacobianM\ into a
product of short-time matrices with matrix elements of comparable magnitudes.
So the problem is reduced to calculating the eigenvalues of the product
of a sequence of matrices.

\subsection{\cLv s}

\emph{multiplicative ergodic theorem}\rf{lyaos,ruelle79} says that the forward and backward
Oseledets matrices
\begin{equation}
\Lambda^{\pm}(x) :=\lim_{t\to\pm\infty}[J^t(x)^\top J^{t}(x)]^{1/2t}
\label{eq:oseledets}
\end{equation}
both exist for an invertible dynamical system equipped with an invariant measure.
The eigenvalues are
$e^{\Lyap^{\pm}_1(x)}<\cdots<e^{\Lyap^{\pm}_s(x)}$, where $\Lyap^{\pm}_i(x)$ are the
Lyapunov exponents (characteristic exponents) and $s$
is the total number of distinct exponents ($s\le n$). For an ergodic system,
Lyapunov exponents are the same almost everywhere, and
$\Lyap^{+}_i(x)=-\Lyap^{-}_{s-i+1}(x)=\Lyap_i$.
The corresponding eigenspaces
$U^\pm_1(x), \cdots, U^\pm_s(x)$
can be used to construct the forward and backward invariant subspaces:
$
V^+_i(x)=U^+_1(x)+\cdots+U^+_i(x)\,,\,
V^-_i(x)=U^-_s(x)+\cdots+U^-_{s-i+1}
$. So the intersections $W_i(x)=V^+_i(x)\cap V^-_i(x)$ are dynamically
forward and backward invariant: $J^{\pm t}(x)W_i(x) \to W_i(f^{\pm t}(x))$,
$i = 1, 2,\cdots,s$.
The expansion rate in invariant subspace $W_i(x)$ is given
by the corresponding Lyapunov exponents,
\begin{equation}
  \label{eq:lyapunov}
  \lim_{t\to\pm\infty}\frac{1}{|t|}\ln\norm{J^t(x)u}
  =\lim_{t\to\pm\infty}\frac{1}{|t|}\ln\norm{[J^t(x)^\top J^t(x)]^{1/2}u}
  = \pm\Lyap_i
  \,,\quad  u\in W_i(x)
\end{equation}
If a Lyapunov exponent has degeneracy one, the corresponding
subspace $W_i(x)$ reduces to a vector, called \emph{covariant vector}.
For \po s, these $\Lyap_i$
(evaluated numerically as $\zeit\to\infty$ limits of many repeats of the
prime period $\period{}$) coincide with the real part of Floquet exponents
(computed in one period of the orbit). Subspace $W_i(x)$ coincides with
a Floquet vector, or, if there is degeneracy, a subspace
spanned by Floquet vectors. 

The reorthonormalization procedure
formulated by Benettin \etc\rf{bene80a}
is 
the standard way to calculate the full spectrum of Lyapunov exponents, 
and it is shown\rf{ErshPot98} 
that the orthogonal vectors produced at the end of 
calculation converges to $U_i^{-}$, eigenvectors of $\Lambda^{-}(x)$, called
the GS vectors (backward Lyapunov vectors). Based on this technique, 
Wolf \etc\rf{WoSa07} and Ginelli \etc\rf{GiChLiPo12} 
invented independent methods to recover \cLv s 
from GS vectors. Here, we should emphasize that GS vectors are 
not invariant. Except the leading one, all of them are dependent on
the specific inner product imposed by the dynamics. Also the local expansion 
rate of \cLv s are not 
identical to the local expansion rate of GS vectors. Specifically for 
\po s, \Fv s depend on no norm, and map forward and
backward as $\jEigvec[j] \to \jMps\,\jEigvec[j]$ under time evolution. 
In contrast, the linearized dynamics does not transport GS vectors into
the tangent space computed further downstream. For more detailed 
comparison, please see\rf{KuPa12, YaRa10}.

\subsection{\CLv s algorithm}
\label{subsec:clvs}
\begin{figure}
  \centering
  \begin{tikzpicture}
  \node (n1) at (0,0) {};
  \node (n2) at (3,-1) {};
  \node (n3) at (7,0.8){};
  \node (n4) at (10,0) {};
  \node (n5) at (11,0){};
  \node (n6) at (12.5,0){};
  
  \draw[very thick, black] (n1) node[left=0.2pt](){$x(t_0)$}
  to[out=0, in=180] (n2) node[fill, red, circle, inner sep=1.5pt](m1){} 
  node[above=0.2pt](){$x(t_1)$}
  to[out=0, in=180] (n3) node[fill, red, circle, inner sep=1.5pt](m2){} 
  node[below=0.2pt](){$x(t_2)$}
  to[out=0, in=180] (n4) node[fill, red, circle, inner sep=1.5pt](m3){} 
  node[below=0.2pt](){$x(t_3)$}
  to[out=0, in=180] (n5);
  
  \draw[very thick, black, ->] (n4) -- (n5);
  
  \draw[very thick, green, decorate,
  decoration={brace,amplitude=5pt,mirror}] 
  ($(n1)+(0.1,-0.3)$) -- ($(n2)+(-0.2,-0.3)$)
  node[black, midway, below=10pt, align=center,rotate=-20]
  {stage 1: $J_iQ_i=Q_{i+1}R_{i}$ \\ forward transient};

  \draw[very thick, green, decorate,
  decoration={brace,amplitude=5pt,mirror}] 
  ($(n2)+(0.1,-0.3)$) -- ($(n4)+(-0.2,-0.3)$)
  node[black, midway, below=10pt, align=center,rotate=10]
  {stage 2: $J_iQ_i=Q_{i+1}R_{i}$ \\ forward, record $Q_i$};

  \draw[very thick, blue, decorate,
  decoration={brace,amplitude=5pt}] 
  ($(n3)+(0.1,0.3)$) -- ($(n4)+(-0.2,0.3)$)
  node[black, midway, above=10pt, align=center,rotate=-15]
  {stage 3: $C_i=R^{-1}_{i}C_{i+1}$ \\ backward transient};

  \draw[very thick, blue, decorate,
  decoration={brace,amplitude=5pt}] 
  ($(n2)+(-0.2,0.4)$) -- ($(n3)+(-0.4,0.4)$)
  node[black, midway, above=10pt, align=center,rotate=23]
  {stage 4: $C_i=R^{-1}_{i}C_{i+1}$ \\backward, record $C_i$};

\end{tikzpicture}
  \caption{Four stages of \CLv s algorithm. The black line is a part of
  a long ergodic trajectory.}
  \label{fig:CLV}
\end{figure}
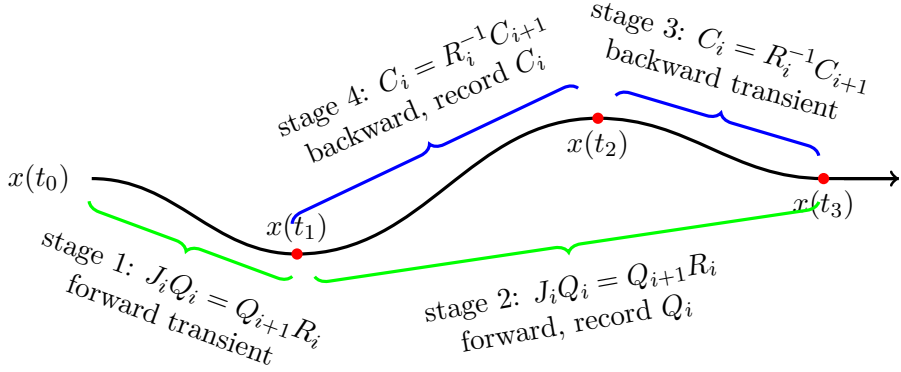
Here we briefly introduce the method used by Ginelli \etc\ to extract 
\cLvs\ from GS vectors. The setup is the same as computing Lyapunov 
exponents. We follow a long ergodic trajectory, and integrate 
the linearized dynamics in tangent space \eqref{eq:tangentDynamics}
with periodic orthonormalization, shown as the first two stages in
\reffig{fig:CLV}. 
Here, $J_i$ is the \JacobianM\ corresponding to
time interval $(t_i, t_{i+1})$, and diagonal elements of upper-triangular 
matrices $R_i$ store local
Lyapunov exponents, long time average of which gives the Lyapunov
exponents of this system. We assume $Q_i$ converges to the GS vectors after
stage 1, and start to record $R_i$ in 
stage 2. Since the first $m$ GS vectors span the same subspace as the 
first $m$ \cLv s, which means
$W_i = Q_i C_i$ \footnote{Here, $W_i$ refers to the matrix formed by individual 
\cLv s at step $i$ of the algorithm. Do not get confused with the $ith$ \cLv 
}
with $C_i$ an upper-triangular matrix, giving the expansion
coefficients of \cLv s in the GS basis.
So we have
$W_i = J_{i-1} Q_{i-1} R^{-1}_{i}C_i = J_{i-1}W_{i-1}C_{i-1}^{-1}R^{-1}_{i} C_i$.
Since $W_i$ is invariant in the tangent space, we must have
$C_{i-1}^{-1}R^{-1}_{i} C_i=I$, which gives the backward dynamics of 
matrix $C_i$ : $C_{i-1} = R^{-1}_{i} C_i$. Ginelli \etc\ cleverly uncover 
this backward dynamics and show that $C_i$ converges after a sufficient
number of iterations (stage 3 in \reffig{fig:CLV}). This process
is continued in stage 4 in \reffig{fig:CLV}, and $C_i$ are recorded 
in this stage. Finally, we obtain the \cLv s for trajectory $x(t_1)$
to $x(t_2)$ in \reffig{fig:CLV}.

\cLv s algorithm is invented to stratify the tangent spaces along
an ergodic trajectory, so it is hard to observe 
degeneracy numerically. However, for periodic orbits, it is 
possible that some Floquet vectors form conjugate complex pairs.
When this algorithm is applied to periodic orbits, it is reduced
to a combination of simultaneous iteration and pure power iteration;
consequently, complex conjugate pairs cannot be told apart.
This 
means that we need to pay attention to the two dimensional rotation 
when checking the convergence of each stages in \reffig{fig:CLV}.
As is shown in latter sections, a complex conjugate pair 
of Floquet vectors can be extracted from this converged two 
dimensional subspace.

\subsection{\Psd\ algorithm}
\label{subsec:psd}
\begin{figure}
  \centering
  \includegraphics[width=0.2\textwidth]{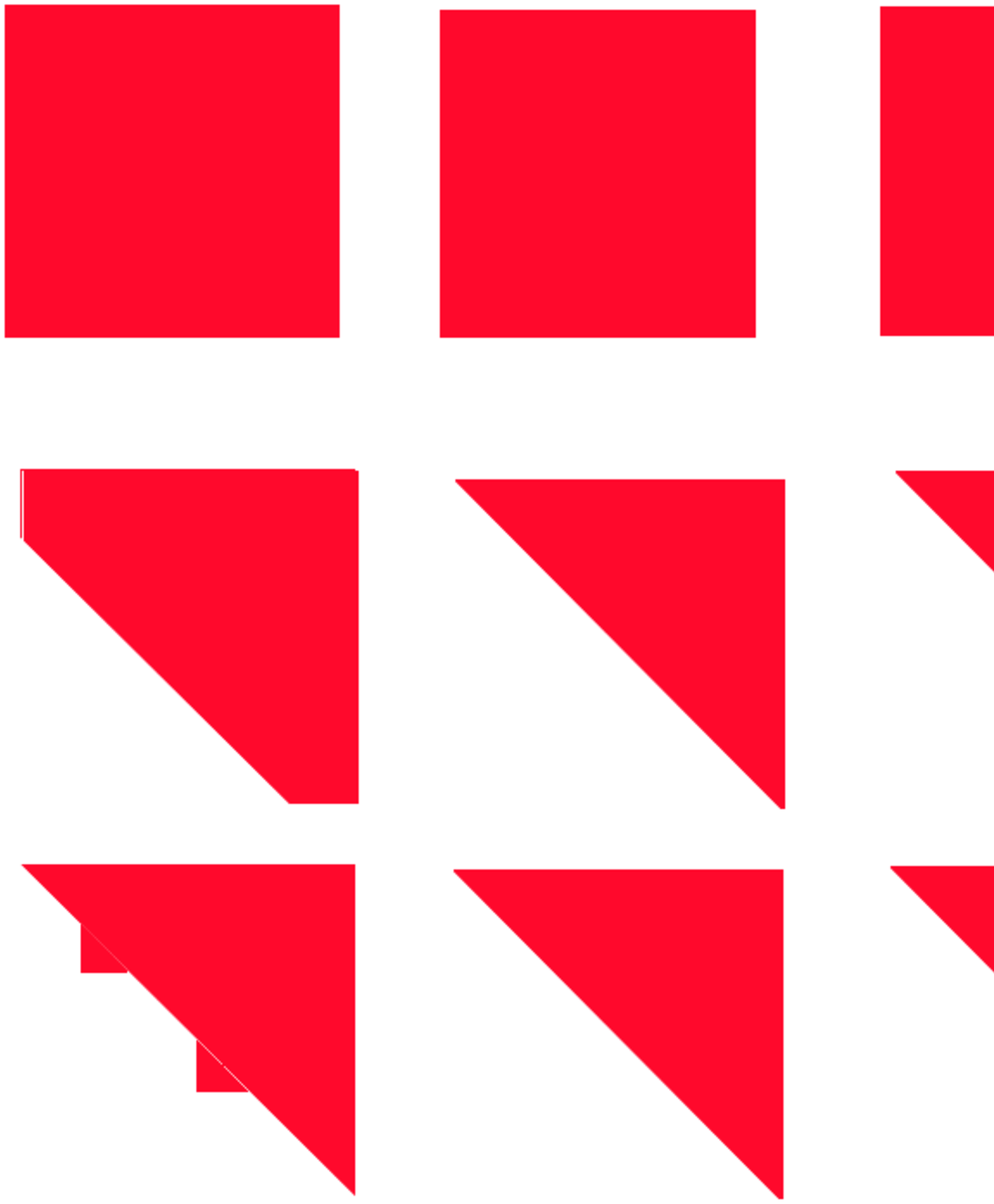}
  \caption{Two stages of \psd\ algorithm illustrated by tree matrices. }
  \label{fig:PSD}
\end{figure}
The implicit QR algorithm (Francis's algorithm) is the standard way 
of solving eigen problem of a single matrix in many numerical packages, such as the
\texttt{eig()} function in Matlab. Bojanczyk \etc\rf{Bojanczyk92theperiodic}
extends the
idea to compute eigenvalues of the product of a sequence of matrices. Later on,
Kurt Lust\rf{Lust01} describes the implementation details and provides
the corresponding 
Fortran code. 
As stated before, by use of chain rule \eqref{eq:xjacobian}, 
\JacobianM\ can be decomposed into a product of short-time
Jacobians with the same dimension, so \psd\ is suitable for computing
Floquet exponents, and we think it is necessary to introduce this algorithm
into physics community.

As illustrated in \reffig{fig:PSD}, \psd\ proceeds in two stages.
First, the sequence
of matrices are transformed to \emph{Hessenberg-Triangular} form,
one of which has upper-Hessenberg form while the others
are upper-triangular
, by a series of Household reflections. The second stage is iteration
of periodic QR algorithm, which diminishes the sub-diagonal components of
the Hessenberg matrix until it becomes quasi-upper-triangular. 
The convergence of this second stage is 
guaranteed by ``Implicit Q Theorem''\rf{Francis61,DSWatkins}. 
After the second stage,
The sequence of matrices are all transformed into upper-triangular form
except one of them be quasi-upper triangular - there are some
$[2\!\times\! 2]$ blocks on the diagonal corresponding to
complex eigenvalues. Then the eigenvalues are
the product of their diagonal elements. 
However, \psd\ is not enough for extracting eigenvectors except the leading one.
Kurt Lust claims to formulate the corresponding \Fv\ algorithm, but to
the best of our knowledge, such algorithm is not present in literature. 
Fortunately, Granat \etc\rf{GranatK06} propose a method to reorder the diagonal elements 
after \psd. It provides a elegant way to compute \Fv s as we will see in
later sections.

\section{Description of the problem}
\label{sect:problem}

After introducing the underlying physical motivation, let us turn
to the definition of the problem. According to
\eqref{eq:xjacobian}, \JacobianM\ can be integrated piece by
piece along a state orbit:
\[
\jMps^{\zeit}(x_0) =
\jMps^{\zeit_{m}-\zeit_{m-1}}(x(\zeit_{m-1}) ,\zeit_{m-1})
\cdots
\jMps^{\zeit_{2}-\zeit_{1}}(x(\zeit_{1}) ,\zeit_{1})
\jMps^{\zeit_{1}-\zeit_{0}}(x(\zeit_{0}) ,\zeit_{0})
\]
with $\zeit_{0}=0$, $\zeit_{m}=t$ and $x_0$ the initial point. For \po s,
$x(\zeit_{m}) = x_0$.
The time sequence $t_i$, $i=1,2,\cdots, m-1$ is
chosen properly such that the elements of each
\JacobianM\ associated with each small time interval has relatively
similar order of magnitude.
For simplicity, we drop all the parameters above and use a bold
letter to denote the product:
\begin{equation}
\ps{\jMps}{0}=\jMps_{m}\jMps_{m-1}\cdots \jMps_{1}\,,\quad
\jMps_{i}\in \mathbb{R}^{n\!\times\! n},\; i\!=\!1,2,\cdots,m
\,.
\label{eq:problem}
\end{equation}
This product can be diagonalized if and only if the sum of
dimensions of eigenspaces of $\ps{\jMps}{0}$ is $n$.
\begin{equation}
  \label{eq:diagonal}
  \ps{\jMps}{0}=E^{(0)}\Sigma(E^{(0)})^{-1}
  \,,
\end{equation}
where $\Sigma$ is a diagonal matrix which stores $\ps{\jMps}{0}$'s
eigenvalues (Floquet multipliers),
$\{ \ExpaEig_{1}, \ExpaEig_{2}, \cdots, \ExpaEig_{n}\}$, and
columns of matrix $E^{(0)}$ are the eigenvectors (Floquet vectors)
of $\ps{\jMps}{0}$:
$E^{(0)}=[\Jve{1}, \Jve{2}, \cdots, \Jve{n}]$. In this paper all
vectors are written in the column form, transpose of $v$ is denoted
$\transp{v}$, and Euclidean `dot' product by $(\transp{v}\,u)$. The
challenge associated with obtaining diagonalized form \eqref{eq:diagonal}
is the fact that often $\ps{\jMps}{0}$ should not be written 
explicitly since the integration process \eqref{eq:tangentDynamics} may overflow
or the resulting matrix is highly ill-conditioned.
Floquet multipliers can easily vary over 100's orders of magnitude,
depending on the system under study and the period of the orbit;
therefore all transformations should be applied to the short time
\JacobianMs\ $J_i$ individually, instead of working with the full-time
$\ps{\jMps}{0}$.
Also, in order to characterize
the geometry along a \po, not only the Floquet
vectors at the initial point are required, but also the
sets at each point on the orbit. Therefore, we also desire
the eigendecomposition of the cyclic rotations of $\ps{\jMps}{0}$:
$\ps{\jMps}{k}=\jMps_{k}\jMps_{k-1}\cdots \jMps_{1}\jMps_{m}\cdots
\jMps_{k+1}$ for $k=1,2,\dots,m\!-\!1$. Eigendecomposition of all
$\ps{\jMps}{k}$ is called the \emph{periodic eigendecomposition} of the
matrix sequence $\jMps_{m}, \jMps_{m-1}, \cdots ,\jMps_{1}$.

The process of implementing eigendecomposition \eqref{eq:diagonal}
proceeds in two stages. First, {\prsf} (PRSF) is
obtained by a similarity transformation for each $\jMps_i$,
\begin{equation}
  \label{eq:prsf}
  \jMps_{i}=Q_{i}R_{i}Q_{i-1}^\top
  \,,
\end{equation}
with $Q_{i}$ orthogonal matrix, and $Q_{0}=Q_{m}$. In the case
considered here, $R_{m}$ is quasi-upper
triangular with $[1\!\times\! 1]$ and $[2\!\times\! 2]$ blocks on the
diagonal, and the remaining $R_{i},i=1,2,\cdots,m\!-\!1$ are upper
triangular. The existence of PRSF, proved in
\refref{Bojanczyk92theperiodic}, provides the \pqr\ that implements \psd.
Defining $\ps{R}{k}=R_{k}R_{k-1}\cdots R_{1}R_{m}\cdots R_{k+1}$, we have
\begin{equation}
  \label{eq:pedrotation}
  \ps{\jMps}{k}=Q_{k}\ps{R}{k}Q_{k}^\top
  \,,
\end{equation}
with the eigenvectors of matrix $\ps{\jMps}{k}$ related to eigenvectors
of quasi-upper triangular matrix $\ps{R}{k}$ by orthogonal matrix
$Q_{k}$. $\ps{\jMps}{k}$ and $\ps{R}{k}$ have the same eigenvalues,
stored in the $[1\!\times\! 1]$ and $[2\!\times\! 2]$ blocks on the
diagonal of $\ps{R}{k}$, and their eigenvectors are transformed by
$Q_{k}$, so the second stage concerns the eigendecomposition of
$\ps{R}{k}$. Eigenvector matrix of $\ps{R}{k}$ has the same structure as
$R_{m}$. We evaluate it by two distinct algorithms. The first one is power iteration
, while the
second algorithm relies on solving a \pse\rf{GranatK06}.

As all $\ps{R}{k}$ have the same eigenvalues, and their eigenvectors are
related by similarity transformations,
\begin{equation}
  \label{eq:Rrelation}
  \ps{R}{k}=(R_{m}\cdots R_{k+1})^{-1}\ps{R}{0}(R_{m}\cdots R_{k+1})
  \,,
\end{equation}
one may be tempted to calculate the eigenvectors of $\ps{R}{0}$, and
obtain the eigenvectors of $\ps{R}{k}$ by \eqref{eq:Rrelation}. The
pitfall of this approach is that numerical errors accumulate when
multiplying a sequence of upper triangular matrices, especially for large
$k$. Therefore, in the second stage of implementing \ped, iteration is
needed for each $\ps{R}{k}$ if power iteration method is chosen in
this stage. \Pse\ bypasses this problem by giving the eigenvectors of all
$\ps{R}{k}$ simultaneously.

Our work illustrates the connection between different algorithms in the
two stages of implementing \ped, pays attention to the case when
eigenvectors appear as complex pairs, and demonstrates that eigenvectors
can be obtained directly from \pse\ without restoring PRSF.

\section{Stage 1 :  periodic real Schur form (PRSF)}
\label{sect:psd}
This is the first stage of implementing \ped.
Eq.~\eqref{eq:pedrotation} represents the eigenvalues of matrix
$\ps{\jMps}{k}$ as real eigenvalues on the diagonal, and complex
eigenvalue pairs as $[2\!\times\! 2]$ blocks on the diagonal of
$\ps{R}{k}$. More specific, if the $i_{th}$
eigenvalue is real, it is given by the product of all the $i_{th}$
diagonal elements of matrices $R_{1},R_{2},\cdots,R_{m}$. In practice,
the logarithms of magnitudes of these numbers are added, in order to
overcome numerical overflows. If the $i_{th}$ and $(i+1)_{th}$
eigenvalues form a complex conjugate pair, all $[2\!\times\! 2]$ matrices
at position $(i,i+1)$ on the diagonal of $R_{1},R_{2},\cdots,R_{m}$ are
multiplied with normalization at each step, and the two complex
eigenvalues of the product are obtained. There is no danger of numerical
overflow because all these $[2\!\times\! 2]$  matrices are in the same
position and  in our applications their elements are of similar order of
magnitude.
Sec.~\ref{subsec:psd} introduce the \psd\ to achieve PRSF. Another 
alternative is the first two stages of \cLv s in sec.~\ref{subsec:clvs},
which reduces to simultaneous iteration for periodic orbits.
Actually, these two methods are equivalent\rf{Trefethen97}, 
but the computational complexity differs.

\paragraph{Simultaneous iteration}

The basic idea of simultaneous iteration is implementing QR decomposition
in the process of power iteration. Assume all Floquet multipliers
are real, without degeneracy, and order them by their
magnitude: $|\ExpaEig_{1}|>|\ExpaEig_{2}|>\cdots >|\ExpaEig_{n}|$, with
corresponding normalized Floquet vectors
$\jEigvec[1], \jEigvec[2],\cdots, \jEigvec[n]$.
For simplicity, here we have dropped the upper indices of these vectors.
An arbitrary initial vector
$\tilde{q}_{1}=\sum_{i=1}^{n}\alpha^{(1)}_{i}\jEigvec[i]$ will converge to the
first Floquet vector $\jEigvec[1]$ after normalization under power iteration of
$\ps{\jMps}{0}$,
\[
 \lim_{\ell\to \infty }\frac{(\ps{\jMps}{0})^{\ell}\tilde{q}_{1}}{||\cdot||}
 \to q_{1}=\jEigvec[1]
 \,.
\]
Here $||\cdot||$ denotes the Euclidean norm of the numerator
($||x||=\sqrt{x^\top x}$). Let $\langle a,b,\cdots,c\rangle$ represent
the space spanned by vector $a,b,\cdots,c$ in $\mathbb{R}^n$. Another
arbitrary vector $\tilde{q}_{2}$ is then chosen orthogonal to subspace
$\langle q_{1} \rangle$ by Gram-Schmidt orthonormalization,
$\tilde{q}_{2}= \sum_{i=2}^{n}\alpha^{(2)}_{i}[\jEigvec[i]-(q_{1}^\top
\jEigvec[i])q_{1}]$.
Note that the index starts from $i=2$ because $\langle q_{1}
\rangle=\langle v_{1} \rangle$. The strategy now is to apply power
iteration of $\ps{\jMps}{0}$ followed by orthonormalization in each
iteration.
\begin{align*}
  \ps{\jMps}{0}\tilde{q}_{2}= &\sum_{i=2}^{n}\alpha^{(2)}_{i}
  [\ExpaEig_{i}\jEigvec[i]-\ExpaEig_{1}(q_{1}^\top \jEigvec[i])q_1] \\
  = & \sum_{i=2}^{n}\alpha^{(2)}_{i}\ExpaEig_{i}[\jEigvec[i]-
  (q_{1}^\top \jEigvec[i])q_{1}]+\sum_{i=2}^{n}\alpha^{(2)}_{i}
  (\ExpaEig_{i}-\ExpaEig_{1})(q_{1}^\top \jEigvec[i])q_{1}
  \,.
\end{align*}
The second term in the above expression will disappear after performing
Gram-Schmidt orthonormalization to $\langle q_{1} \rangle$, and the first
term will converge to
$q_{2}=\jEigvec[2]-(q_{1}^\top \jEigvec[2])q_{1}$ (not
normalized) after a sufficient number of iterations because of the
decreasing magnitudes of $\ExpaEig_{i}$, and we also note that $\langle
v_{1}, v_{2}\rangle=\langle q_{1}, q_{2}\rangle$. The same argument can
be applied to $\tilde{q}_{i},\;i=3,4,\cdots,n$ as well.
In this way, after a sufficient number of iterations,
\[
\lim_{\ell\to \infty}(\ps{\jMps}{0})^{\ell}[\tilde{q}_{1},\tilde{q}_{2},\cdots,
\tilde{q}_{n}]
\to [q_{1},q_{2}\cdots, q_{n}]
\:,
\]
where
\[
\begin{aligned}
  & q_{1} = \jEigvec[1]\,,\qquad
    q_{2} = \frac{\jEigvec[2]-(\jEigvec[2]^\top q_{1})q_{1}}{||\cdot ||}\,,
    \quad \cdots\,,\quad
    q_{n} = \frac{\jEigvec[n]-\sum_{i=1}^{n-1}(\jEigvec[n]^\top
      q_{i})q_{i}}{||\cdot ||}
\,.
\end{aligned}
\]
Let matrix $Q_{0}=[q_{1},q_{2},\cdots ,q_{n}]$; then we have
$\ps{\jMps}{0}Q_{0}=Q_{0}\ps{R}{0}$ with $\ps{R}{0}$ an upper triangular
matrix because of $\langle q_{1},q_{2},\cdots,q_{i} \rangle=\langle
v_{1},v_{2},\cdots,v_{i} \rangle$, which is just
$\ps{\jMps}{0}=Q_{0}\ps{R}{0}Q^\top_{0}$ (the Schur decomposition of
$\ps{\jMps}{0}$). The diagonal elements of $\ps{R}{0}$ are the
eigenvalues of $\ps{\jMps}{0}$ in decreasing order.
Numerically, the process described above can be implemented on an
arbitrary initial full rank matrix $\tilde{Q}_0$ followed by QR
decomposition at
each step $\jMps_{s}\tilde{Q}_{s-1}=\tilde{Q}_{s}\tilde{R}_{s}$ with
$s=1,2,3,\cdots$ and $\jMps_{s+m}=\jMps_{s}$. For sufficient number of
iterations, $\tilde{Q}_{s}$ and $\tilde{R}_{s}$ converge to $Q_{s}$ and
$R_{s}$ \eqref{eq:prsf} for $s=1,2,\cdots,n$, so we achieve
\eqref{eq:pedrotation} the \psd\ of $\ps{\jMps}{k}$.

We have thus demonstrated that simultaneous iteration converges to
PRSF for real non-degenerate eigenvalues.
For complex eigenvalue pairs, the algorithm converges in the sense that
the subspace spanned by a complex conjugate vector pair converges. So,
\[
\ps{\jMps}{0}Q_{0}=Q^{'}_{0}\ps{R}{0}=Q_{0}D\ps{R}{0}
\,,
\]
where $D$ is a block-diagonal matrix with diagonal elements $\pm 1$
(corresponding to real eigenvalues) or $[2\!\times\! 2]$ blocks
(corresponding to complex eigenvalue pairs). Absorb $D$ into $R_{m}$,
then $R_{m}$ becomes a quasi-upper triangular matrix, and \eqref{eq:prsf}
still holds.

\section{Stage 2 : eigenvector algorithms}
\label{sect:eigenvec}

Upon achieving PRSF,
the eigenvectors of $\ps{\jMps}{k}$ are related to eigenvectors of
$\ps{R}{k}$ by orthogonal matrix $Q_{k}$ from \eqref{eq:prsf}, and
the
eigenvector matrix of $\ps{R}{k}$ has the same quasi-upper triangular
structure as $R_m$. In addition, if we follow the
simultaneous iteration method or
implement \psd\ without shift, eigenvalues are ordered by their
magnitudes on the diagonal. Power iteration utilizing this property
could be easily implemented to generate the eigenvector
matrix. This is the basic idea of the first algorithm for generating
eigenvectors of $\ps{R}{k}$, corresponding to the 3rd and 4th stage in
\cLv s algorithm in figure \reffig{fig:CLV}.
Alternatively,
observation that
the first eigenvector of $\ps{R}{k}$ is trivial if it is real,
$\Rve{1}=(1,0,\cdots,0)^\top $, inspires us to reorder the
eigenvalues so that the $j_{th}$ eigenvalue is in the first diagonal
place of $\ps{R}{k}$; in this way, the $j_{th}$ eigenvector is obtained.
For both methods, attention should be paid to the complex conjugate
eigenvector pairs. In this section, $\Rve{i}^{(k)}$
denotes the $i_{th}$ eigenvectors of $\ps{R}{k}$, contrast to
$\jEigvec[i]^{(k)}$
the eigenvectors of $\ps{J}{k}$, and for most cases, the upper indices
are dropped if no confusion occurs.

\subsection{Iteration method}

The prerequisite for iteration method is that all the eigenvalues are
ordered in a ascending or descending way by their magnitude on the
diagonal of $\ps{R}{k}$. Assume that they are in descending order, which
is the outcome of simultaneous iteration; therefore the diagonal elements
of $\ps{R}{k}$ are $\ExpaEig_{1},\ExpaEig_{2},\cdots,\ExpaEig_{n}$, with
magnitudes from large to small.
If the $i_{th}$ eigenvector of $\ps{R}{k}$ is real, then it has form
$\Rve{i}=(a_{1},a_{2},\allowbreak \cdots,a_{i},0,\cdots, 0)^\top $. An arbitrary
vector whose first $i$ elements are nonzero
$x=(b_{1},b_{2},\cdots,b_{i},0, \allowbreak \cdots, 0)^\top $ is a linear combination
of the first $i$ eigenvectors: $x=\sum_{j=1}^{i}\alpha_{j}\Rve{j}$.
Use it as the initial condition for the power iteration by
$(\ps{R}{k})^{-1}=R_{k+1}^{-1}\cdots R_{m}^{-1}R_{1}^{-1}R_{2}^{-1}\cdots
R_{k}^{-1}$ and after sufficient number of iterations,
\[
\lim_{\ell\to \infty} \frac{(\ps{R}{k})^{-\ell}x}{||\cdot||}=\Rve{i}
\,.
\]
The property we used here is that $(\ps{R}{k})^{-1}$ and $\ps{R}{k}$ have
the same eigenvectors but inverse eigenvalues.

For a $[2\!\times\! 2]$ block on the diagonal of $\ps{R}{k}$, the
corresponding conjugate complex eigenvectors form a two dimensional subspace.
Any real vector selected from this subspace will rotate under power 
iteration. In this case, power iteration still converges in the sense 
that the subspace spanned by the complex
conjugate eigenvector pair converges.
Suppose the $i_{th}$ and $(i+1)_{th}$ eigenvectors of $\ps{R}{k}$ form a
complex pair. Two arbitrary vectors $x_{1}$ and $x_{2}$ whose first $i+1$
elements are non zero can be written as the linear superposition of the
first $i+1$ eigenvectors,
$x_{1,2}=(\sum_{j=1}^{i-1}\alpha^{(1,2)}_{j}\Rve{j})+\alpha^{(1,2)}_{i}\Rve{i}+
(\alpha^{(1,2)}_{i}\Rve{i})^{*}
$,
where $(*)$ denotes the complex conjugate. As for the real case, the
first $i\!-\!1$ components above will vanish after a sufficient number of
iterations. Denote the two vectors at this instance (corresponding to
$x_{1,2}$) to be $X_{1}$ and $X_{2}$ and form matrix $X=[X_{1},X_{2}]$.
The subspace spanned by $X_{1,2}$ does not change and $X$ will be rotated
after another iteration,
\begin{equation}
(\ps{R}{k})^{-1}X=X^{'}=XC
\,,
\label{eq:similar}
\end{equation}
where $C$ is a $[2\!\times\! 2]$ matrix which has two complex conjugate
eigenvectors $\Rve{C}$ and $(\Rve{C})^{*}$. Transformation
\eqref{eq:similar} relates the eigenvectors of $\ps{R}{k}$ with those of
$C$: $[\Rve{i},(\Rve{i})^{*}]=X[\Rve{C},(\Rve{C})^{*}]$.
In practice, matrix $C$ can be computed by QR decomposition; let
$X=Q_{X}R_{X}$ be the QR decomposition of $X$, then
$C=R_{X}^{-1}Q_{X}^\top X^{'}$.
On the other hand,
complex eigenvectors are not uniquely determined in the sense that
$e^{i\theta}\Rve{i}$ is also a eigenvector with the same eigenvalue
as $\Rve{i}$ for an arbitrary angle $\theta$, so when comparing
results from different eigenvector algorithms, we need a constraint to
fix the phase of a complex eigenvector, such as letting the first element
be real.

We should note that performance of power iteration depends on the ratios
of magnitudes of eigenvalues, so performance is poor for systems with
clustered eigenvalues. We assume that proper modifications, such as 
shifted iteration or inverse iteration, may help improve the performance.
Such techniques are beyond the scope of this paper.

\subsection{reordering method}
\label{sect:reorder}
Except for the performance problem with clustered eigenvalues, the 
power iteration has a more severe issue when applied to dynamical systems, 
that is, it cannot get the eigenvectors of $\ps{R}{k}$ for all $k\in
{0,1,2,\cdots,m}$ at the same time. Although eigenvectors of $\ps{R}{k}$
and $\ps{R}{0}$ are related by \eqref{eq:Rrelation}, it is not advisable,
as pointed out above, to evolve the eigenvectors of $\ps{R}{0}$ so as to
get eigenvectors of $\ps{R}{k}$ because of the noise introduced during
this process. Therefore, iteration is needed for each $k\in
{0,1,2,\cdots,m}$.

There exists a direct algorithm to obtain the eigenvectors of every
$\ps{R}{k}$ at once without iteration. The idea is very simple: the
eigenvector corresponding to the first diagonal element of an
upper-triangular matrix is $\Rve{1}=(1,0,\cdots,0)^\top $. By
reordering the diagonal elements (or $[2\!\times\! 2]$ blocks) of
$\ps{R}{0}$, we can find any eigenvector by positioning the corresponding
eigenvalue in the first diagonal position. Although in our application
only reordering of $[1\!\times\! 1]$ and $[2\!\times\! 2]$ blocks is
needed, we recapitulate here the general case of reordering two adjacent
blocks of a quasi-upper triangular matrix following Granat\rf{GranatK06}.
Partition $R_{i}$ as
\[
R_{i}=
\left[
\begin{array}{c|cc|c}
  R^{00}_{i} & * & *& * \\ \hline
  0 & R^{11}_{i} & R^{12}_{i} & * \\
  0 & 0 & R^{22}_{i} & * \\ \hline
  0 & 0 & 0 & R^{33}_{i}
\end{array}
\right]
\,,
\]
where $R^{00}_{i}, R^{11}_{i},R^{22}_{i},R^{33}_{i}$ have size
$[p_{0}\!\times\! p_{0}], [p_{1}\!\times\! p_{1}], [p_{2}\!\times\!
p_{2}]$ and $[p_{3}\!\times\! p_{3}]$ respectively, and
$p_{0}+p_{1}+p_{2}+p_{3}=n$. In order to exchange the middle two blocks
($R^{11}_{i}$ and $R^{22}_{i}$), we construct a non-singular periodic
matrix sequence: $\hat{S_{i}},\:i=0,1,2,\cdots,m$ with
$\hat{S_{0}}=\hat{S_{m}}$,
\[
\hat{S_{i}}=
\left[
\begin{array}{c|c|c}
  I_{p_{0}} & 0 & 0  \\ \hline
  0 & S_{i} & 0 \\ \hline
  0 & 0 & I_{p_{3}}
\end{array}
\right]
\,,
\]
where $S_{i}$ is a $[(p_{1}+p_{2})\!\times\! (p_{1}+p_{2})]$ matrix,
such that $\hat{S}_{i}$ transforms $R_{i}$ as follows:
\begin{equation}
\label{eq:xdtransform}
\hat{S}_{i}^{-1}R_{i}\hat{S}_{i-1}=\tilde{R}_{i}=
\left[
\begin{array}{c|cc|c}
  R^{00}_{i} & * & *& * \\ \hline
  0 & R^{22}_{i} & 0 & * \\
  0 & 0 & R^{11}_{i} & * \\ \hline
  0 & 0 & 0 & R^{33}_{i}
\end{array}
\right]
\,,
\end{equation}
which is
\[
S^{-1}_{i}
\left[
\begin{array}{c c}
  R^{11}_{i} & R^{12}_{i} \\
  0 & R^{22}_{i}
\end{array}
\right]
S_{i-1}=
\left[
\begin{array}{c c}
  R^{22}_{i} & 0 \\
  0 & R^{11}_{i}
\end{array}
\right]
\,.
\]
The problem is to find the appropriate matrix $S_{i}$ which satisfies
the above condition. Assume $S_{i}$ has form
\[
S_{i}=
\left[
\begin{array}{c c}
  X_{i} & I_{p_{1}} \\
  I_{p_{2}} & 0
\end{array}
\right]
\,,
\]
where matrix $X_{i}$ has dimension $[p_{1}\!\times\! p_{2}]$. We obtain
periodic Sylvester equation\rf{GranatK06}
\begin{equation}
  \label{eq:xdpse}
  R^{11}_{i}X_{i-1}-X_{i}R^{22}_{i}=-R^{12}_{i}
  \,,\quad i=0,1,2,\cdots,m
  \,.
\end{equation}

The algorithm to find eigenvectors is based on \eqref{eq:xdpse}. If the
$i_{th}$ eigenvalue of $\ps{R}{k}$ is real, we only need to exchange the
first $[(i-1)\!\times\! (i-1)]$ block of $R_{k}\,,k=1,2,\cdots,m$ with
its $i_{th}$ diagonal element. If the $i_{th}$ and $(i+1)_{th}$
eigenvalues form a complex pair, then the first $[(i-1)\!\times\! (i-1)]$
block and the following $[2\!\times\! 2]$ block should be exchanged.
Therefore $X_{i}$ in \eqref{eq:xdpse} has dimension $[p_{1}\!\times\! 1]$
or $[p_{1}\!\times\! 2]$. In both cases, $p_{0}=0$.

\paragraph{Real eigenvectors}
In this case, matrix $X_{i}$ is just a column vector, so
\eqref{eq:xdpse} is equivalent to
\begin{equation}
  \label{eq:xdpsereal}
  \begin{bmatrix}
    R^{11}_{1} & -R^{22}_{1}I_{p_{1}} &  & \\[1em]
    & R^{11}_{2} & -R^{22}_{2}I_{p_{1}} &  &\\[1em]
    &  & R^{11}_{3} & -R^{22}_{3}I_{p_{1}} &  &\\[1em]
    & & & \ddots &\cdots & \\[1em]
    -R^{22}_{m}I_{p_{1}} & & & & R^{11}_{m}
  \end{bmatrix}
  \begin{bmatrix}
    X_{0} \\[1em]
    X_{1}  \\[1em]
    X_{2}  \\[1em]
    \cdots \\[1em]
    X_{m-1}
  \end{bmatrix}
  =
  \begin{bmatrix}
    -R^{12}_{1} \\[1em]
    -R^{12}_{2} \\[1em]
    -R^{12}_{3} \\[1em]
    \cdots \\[1em]
    -R^{12}_{m}
  \end{bmatrix}
\,,
\end{equation}
where $R^{22}_{i}$ is the $(p_{1}+1)_{th}$ diagonal element of $R_{i}$. 
The accuracy of eigenvectors is determined by the accuracy of
solving sparse linear equation \eqref{eq:xdpsereal}. In our application
to periodic orbits in one dimensional \KSe, 
Gaussian  elimination with partial pivoting (GEPP) is enough. For a more
technical treatment, such as cyclic reduction or preconditioned conjugate
gradients, to name a few, please see\rf{NLA:NLA198,aabdls,GranatRBA}.

Now we get all vectors $X_{i}$ by solving \pse, but how are they related
to the eigenvectors? In analogy to $\ps{R}{0}$, defining
$\mathbf{\tilde{R}}_{0}=\tilde{R}_{m}\tilde{R}_{m-1}\cdots
\tilde{R}_{1}$, we get
$\hat{S}_{m}^{-1}\ps{R}{0}\hat{S}_{m}=\mathbf{\tilde{R}}_{0}$ by
\eqref{eq:xdtransform}. Since $p_{0}=0$ and $p_{2}=1$ in
\eqref{eq:xdtransform}, the first eigenvector of
$\mathbf{\tilde{R}}_{0}$, the one corresponding to eigenvalue
$\ExpaEig_{p_1+1}$
is $\tilde{e}=(1,0,\cdots , 0)^\top $. Before
normalization, the corresponding eigenvector of $\ps{R}{0}$ is
\[
\Rve{p_{1}+1}^{(0)}=\hat{S}_{m}\tilde{e}
 = \left[X_{0}^\top , 1, 0, 0, \cdots, 0 \right]^\top
\,.
\]
This is the eigenvector of matrix $\ps{R}{0}=R_{m}R_{m-1}\cdots R_{1}$ in
\eqref{eq:pedrotation} for $k=0$. For $\ps{R}{1}=R_{1}R_{m}\cdots R_{2}$,
the corresponding \pse\ will be cyclically rotated one row up, which
means $X_{1}$ will be shifted to the first place in the column vector in
\eqref{eq:xdpsereal}, and thus the corresponding eigenvector of
$\ps{R}{1}$ is $\Rve{p_{1}+1}^{(1)}=[X_{1}^\top,1,0,\cdots,0]^\top $. The
same argument goes for all the following $\ps{R}{k}\,,k=2,3,\cdots,m-1$.
In conclusion, solution of \eqref{eq:xdpsereal} contains the eigenvectors
for all $\ps{R}{k}\,,k=0,1,\cdots,m-1$.
Another benefit of reordering method is that we can selectively
get the eigenvectors corresponding to some specific eigenvalues.
This merit is important in high dimensional nonlinear systems for
which only a subset of Floquet vectors suffices to characterize the
dynamics in tangent space, and thus we avoid wasting time in calculating
the remaining transient uniformly vanishing modes.

\paragraph{Complex eigenvector pairs}
As in the real eigenvalue case, we have $p_{0}=0$, but now $p_{2}=2$, so
matrix $X_{i}$  has dimension $[p_{1}\!\times\! 2]$. Using the same
notation as \refref{GranatK06}, let $v(X_{i})$ denote the vector
representation of $X_{i}$ with the columns of $X_{i}$ stacked on top of
each other, and let $A\otimes B$ denote the Kronecker product of two
matrices, with the $(i,j)$-block element be $a_{ij}B$.

Now, the \pse\ \eqref{eq:xdpse} is equivalent to
\begin{equation}
  \label{eq:xdpsdcomplex}
  \resizebox{\linewidth}{!}{%
  $
  \setlength{\arraycolsep}{3pt}
  \begin{bmatrix}
    I_{2}\otimes R^{11}_{1} & -(R^{22}_{1})^\top \otimes I_{p_{1}} &  & \\[1em]
    & I_{2}\otimes R^{11}_{2} & -(R^{22}_{2})^\top  \otimes I_{p_{1}} &  &\\[1em]
    &  & I_{2}\otimes R^{11}_{3} & -(R^{22}_{3})^\top \otimes I_{p_{1}} &  &\\[1em]
    & & & \ddots &\cdots & \\[1em]
    -(R^{22}_{m})^\top \otimes I_{p_{1}} & & & & I_{2}\otimes R^{11}_{m}
  \end{bmatrix}
  \begin{bmatrix}
    v(X_{0}) \\[1em]
    v(X_{1})  \\[1em]
    v(X_{2})  \\[1em]
    \cdots \\[1em]
    v(X_{m-1})
  \end{bmatrix}
  =
  \begin{bmatrix}
    -v(R^{12}_{1}) \\[1em]
    -v(R^{12}_{2}) \\[1em]
    -v(R^{12}_{3}) \\[1em]
    \cdots \\[1em]
    -v(R^{12}_{m})
  \end{bmatrix} $%
}
\,.
\end{equation}
After switching $R^{11}_{i}$ and $R^{22}_{i}$, we can get the first two
eigenvectors of $\mathbf{\tilde{R}}_{0}$ by multiplying the first
$[2\!\times\! 2]$ diagonal blocks of $\tilde{R_{i}}$:
$R^{22}=R^{22}_{m}R^{22}_{m-1}\cdots R^{22}_{1}$. Let the eigenvectors of
$R^{22}$ be $v$ and $v^{*}$ of size $[2\!\times\! 1]$, then the
corresponding eigenvectors of $\mathbf{\tilde{R}}_{0}$ are
$\tilde{e}_{1}=(v^\top,0,0,\cdots,0)^\top $ and
$\tilde{e}_{2}=(\tilde{e}_{1})^{*}$ (the additional zeros make the length
of the eigenvectors to be $n$). Therefore, the corresponding eigenvectors
of $\ps{R}{0}$ are
\[
\left[\Rve{p_{1}+1}^{(0)},\Rve{p_{1}+2}^{(0)}\right]
 =\hat{S}_{m}[\tilde{e}_{1},\tilde{e}_{2}]
 = \left[
   \begin{array}{c}
   X_{0} \\
   I_{2}    \\
   0 \quad 0\\
   0 \quad 0\\
   \vdots\\
   0 \quad 0
   \end{array}
 \right]
 [v,v^{*}]
\,.
\]
For other $\ps{R}{k}$, the same argument in the real case applies
here too, so we obtain all the complex eigenvector pairs for
$\ps{R}{k}\,,k=1,2,\cdots,m$.

\section{Computational complexity and convergence analysis}
\label{sect:error}

In this paper we make no attempt at conducting a strict error analysis of
the algorithms presented. However, for practical applications
it is important to understand their computational costs.
\Ped\ is conducted in two stages: (1) {\prsf},
and (2) determination of all
eigenvectors. In each stage, there are two candidate algorithms, so the
efficiency of \ped\ depends on the choice of the specific algorithm
chosen in each stage.

\Pqr\ and simultaneous iteration are both effective to achieve PRSF for
the real eigenvalues, and for complex pairs of eigenvalues. \Pqr\
consists of two stages. First, matrix sequence
$\jMps_{m},\jMps_{m-1}\cdots,\jMps_{1}$ is reduced to
Hessenberg-triangular form, with $\jMps_{m-1},\cdots,\jMps_{1}$ upper
triangular and $\jMps_{m}$ upper Hessenberg. It requires $O(mn)$
Householder reflections in this stage and computational cost associated
with each reflection is $O(n^{2})$, if the transformed matrix is
calculated implicitly without forming the Householder
matrix\rf{Trefethen97}. So the overall computational cost of this stage
is $O(mn^{3})$. The second stage is the periodic QR iteration which is a
generalization of the standard, $m=1$, case\rf{Trefethen97}. $O(mn)$
Givens rotations are performed in each iteration with overall
computational cost of $O(mn^{2})$. Though the computational effort in
each iteration in the second stage is less than that in the first stage,
the number of iterations in the second stage is usually far more than the
dimension of matrices involved. In this sense, the second stage is the
heavy part of \pqr. On the other hand, simultaneous iteration conducts
$m$ QR decomposition $O(mn^{3})$ and $m$ matrix-matrix multiplication
$O(mn^{3})$ in each iteration, giving a total computational cost of
$O(mn^{3})$. The convergence of either algorithm depends
linearly on the ratio of adjacent eigenvalues of $\ps{R}{0}$:
$|\ExpaEig_{i}|/|\ExpaEig_{i+1}|$  without shift\rf{Francis61}. Therefore
the ratio of costs is approximately of the order $O(mn^3)/O(mn^2) = O(n)$,
implying that the periodic QR algorithm is much cheaper than the
simultaneous iteration if the dimension of matrices involved is large
enough.

The second stage of \ped\ is to find all the eigenvectors of
$\ps{\jMps}{k}$ via quasi-upper triangular matrices $\ps{R}{k}$. The
first candidate is the combination of power iteration and shifted power
iteration. The computational cost of one iteration for the $i_{th}$
eigenvector is $O(mi^{2})$. The second candidate, reordering method,
relies on an effective method to solve \pse\ \eqref{eq:xdpse}. For
example, GEPP is suitable
for well conditioned matrix \eqref{eq:xdpsereal} and
\eqref{eq:xdpsdcomplex} with computational cost of $O(mn^{2})$. On the
other hand, the iteration method, as pointed out earlier, could not
produce the eigenvectors of $\ps{R}{k}$ for all $k=1,2,\cdots,m$
accurately in the same time due to the noise introduced during the
transformation process \eqref{eq:pedrotation}, especially when the
magnitudes of eigenvalues span a large range. In contrast, the reordering
algorithm is not iterative and it gives all the eigenvectors
simultaneously.

In summary, if we just consider the computational effort, the combination
of \pqr\ and reordering method is preferable for \ped.

\section{Application to \KSe}
\label{sect:applic}
Our ultimate goal of implementing \ped\ is to analyze the stability 
of periodic orbits and the associated stable/unstable manifolds in
dynamical systems, for the hope of getting a better understanding 
of pattern formation and turbulence. 
As an example, we focus on the one-dimensional \KSe\
\begin{equation}
u_t+\frac{1}{2}(u^2)_x+u_{xx}+u_{xxxx}=0\,,\; x\in [0,L]
\label{eq:ks}
\end{equation}
on a periodic spatial domain of size $L = 22$, large enough to exhibit
complex spatiotemporal dynamics. This equation is formulated 
independently by Kuramoto in the context of angular phase 
turbulence in reaction-diffusion systems\rf{kuramoto1975}, and
by Sivashinsky in the study of hydrodynamic instability in laminar 
flames\rf{michsiv77}. 
Periodic
boundary condition enables us to transform this partial differential
equation into a set of ODEs in Fourier space
\begin{equation}
\dot{a}_k \;\; =
( q_k^2 - q_k^4 )\, a_k
- i \frac{q_k}{2} \sum_{m=-\infty}^{\infty}a_m a_{k-m}
\label{eq:ksfourier} 
\end{equation}
where $q_k = 2\pi k/L$, and the coefficients are complex,
$a_{k}=b_{k}+ic_{k}$. In our simulations, discrete Fourier transform 
is used with $N=64$ modes ($k = -N/2 + 1$ up to $N/2$ in \eqref{eq:ksfourier}).

Since $u(x,t)$ is real, $a_{k}(t)=a^{*}_{-k}(t)$; thus only half of the
Fourier modes are independent. As $\dot{a}_{0}=0$ from
\eqref{eq:ksfourier}, we can set $a_{0}=0$ corresponding to 
zero mean velocity without lose of generality. 
Also the nonlinear term 
of $\dot{a}_{N/2}$ in fact has coefficient 
$q_{N/2} + q_{-N/2} = 0$ from symmetric consideration\rf{trefethenSpectral};
thus $a_{N/2}$ is decoupled from other modes and it
can be set to zero as well. Thus then the number of independent variables
is $N-2$,
\begin{equation}
\label{eq:fourierspace}
\hat{u}=(b_{1},c_{1},b_{2},c_{2},\cdots,b_{N/2-1},c_{N/2-1})^\top
\,.
\end{equation}
This is the `\statesp' in the discussion that follows. Exponential time-differencing
scheme combined with RK4\rf{cox02jcomp, ks04com} is implemented to integrate
\eqref{eq:ksfourier}. The combination of \pqr\ algorithm and reordering
algorithm is used to obtain all exponents and eigenvectors. In addition,
Gaussian  elimination with partial pivoting (GEPP) is stable for
\eqref{eq:xdpsereal} and \eqref{eq:xdpsdcomplex} if the
time step in \KS\ integrator is not too
large, as GEPP only uses addition and subtraction operations.
\begin{figure}[h]
  \centering
  \begin{minipage}{.22\textwidth}
    \centering \small{\texttt{(a)}}
    \includegraphics[width=\textwidth]{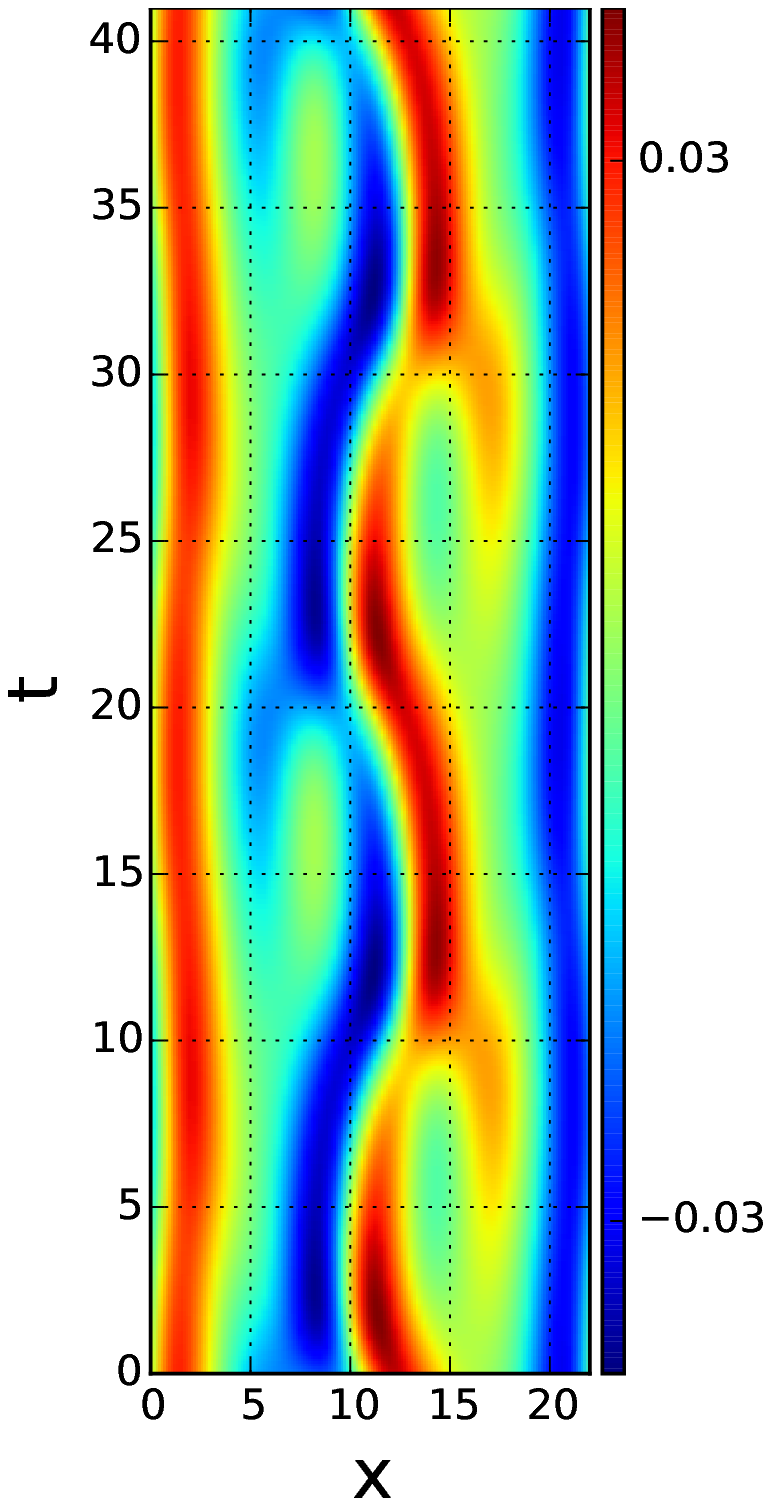}
  \end{minipage}
  \begin{minipage}{.22\textwidth}
    \centering \small{\texttt{(b)}}
    \includegraphics[width=\textwidth]{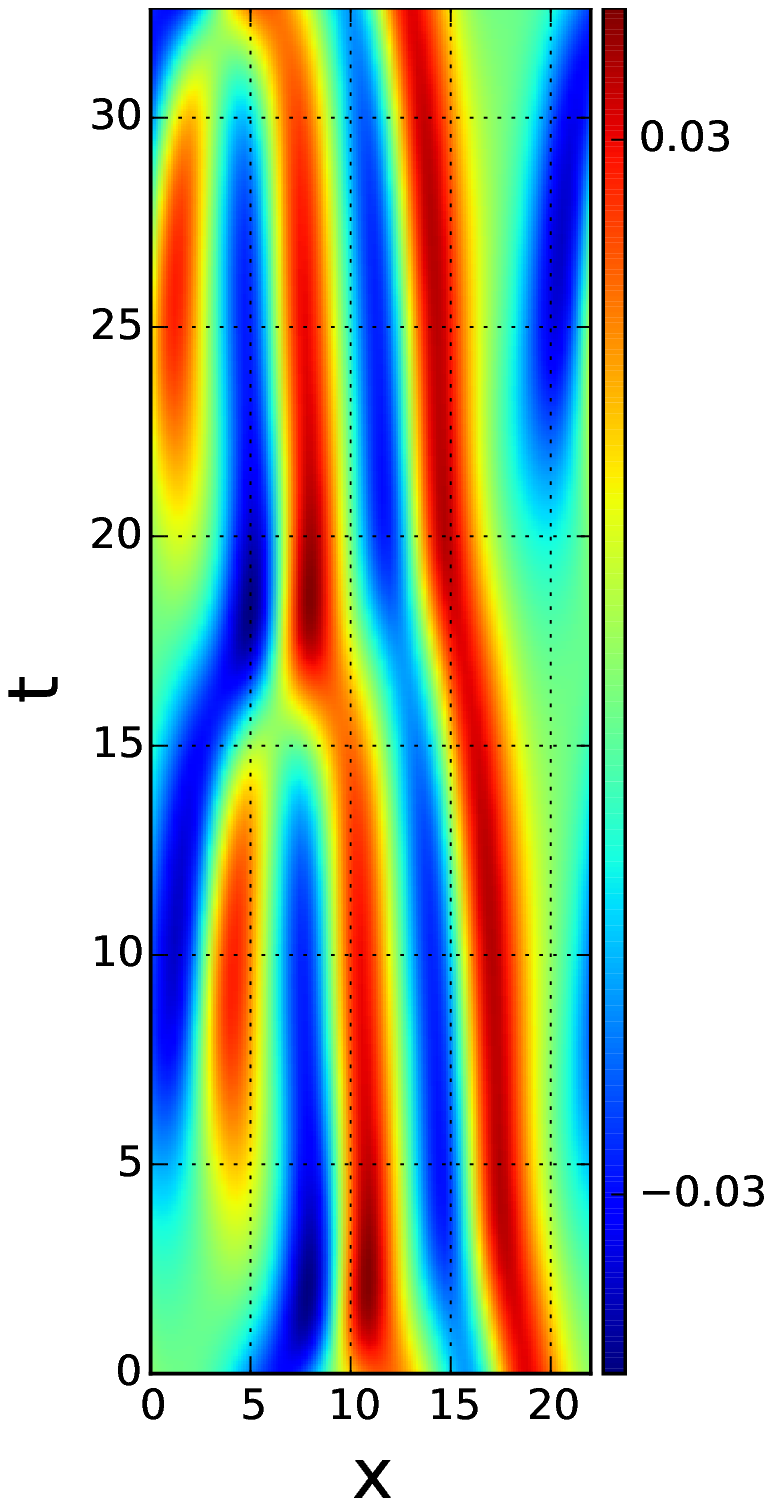}
  \end{minipage}%
  \begin{minipage}{.55\textwidth}
    \centering \small{\texttt{(c)}}
    \includegraphics[width=\textwidth]{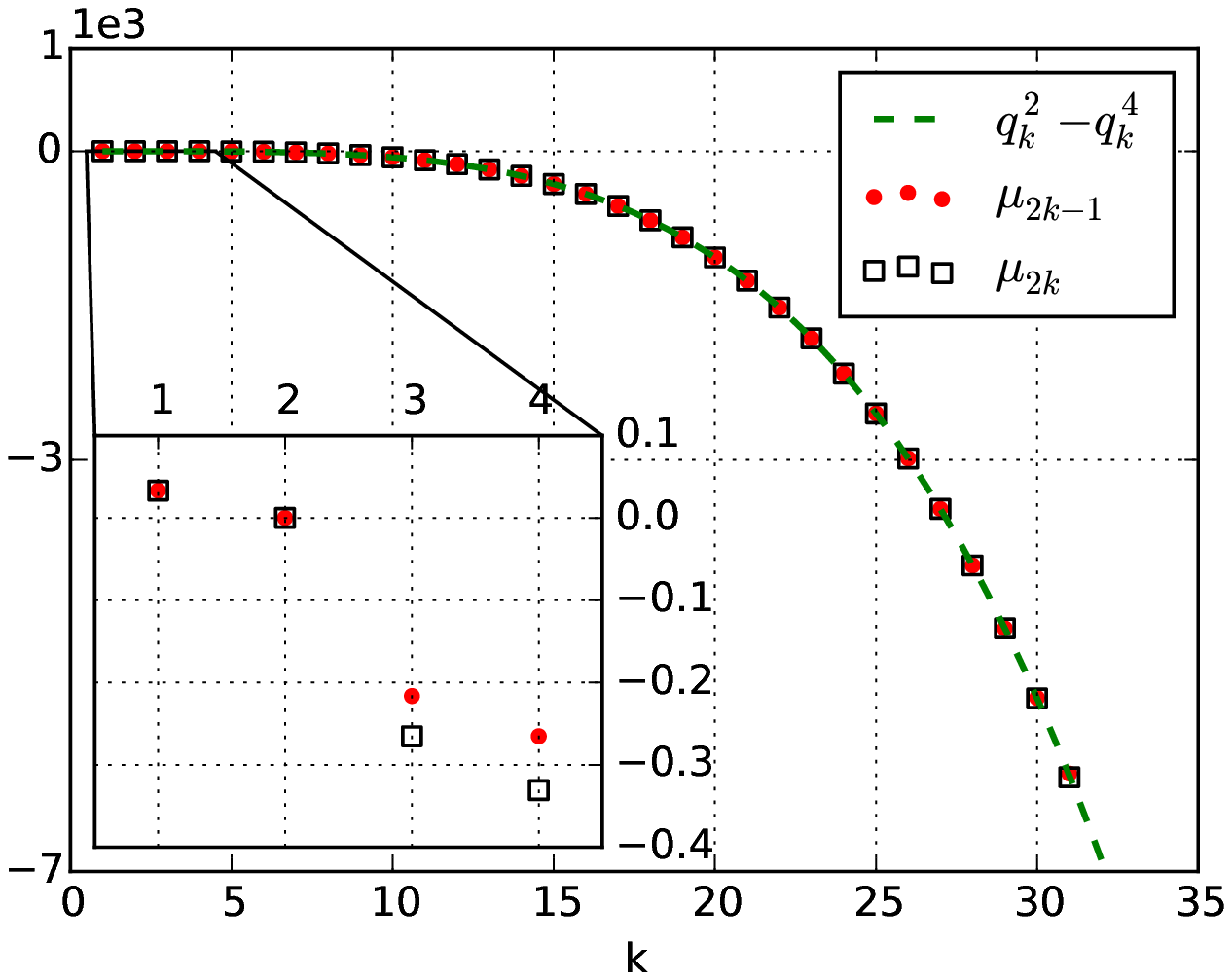}
  \end{minipage}
  \caption{(Color online)
   (a) Pre\po\ $\cycle{pp}_{10.25}$ and
   (b) \rpo\ $\cycle{rp}_{16.31}$ for total evolution time
   $4\,\period{pp}$ and $2\,\period{rp}$, respectively. The phase shift
   for $\cycle{rp}_{16.31}$ after one prime period $\simeq-2.863$.
   (c) The real parts of Floquet exponents paired for a given $k$ as
   $(k,\eigRe[2k-1])$ and $(k,\eigRe[2k])$, for $\cycle{pp}_{10.25}$ with
   truncation number $N=64$. The dashed line (green) is
   $q_{k}^{2}-q_{k}^{4}$. The inset is a magnification of the region
   containing the 8 leading {\entangled} modes. 
  }
  \label{fig:ppo1rpo1}
\end{figure}
\begin{table}[h]
  \footnotesize
  \centering
  \caption{
    The first 10 and last four Floquet exponents and
    Floquet multiplier phases,
    $ \ExpaEig_i= \exp(\period{}\,\eigRe[i] \pm i\theta_{i})$, for
    orbits $\cycle{pp}_{10.25}$ and $\cycle{rp}_{16.31}$, respectively.
    $\theta_{i}$ column lists either the phase,
    if the Floquet multiplier is complex, or `-1' if the
    multiplier is real, but inverse hyperbolic. Truncation number
    $N=64$.
    The $8$ leading exponents correspond to the {\entangled} modes:
    note the sharp drop in the value of the $9_{th}$ and subsequent
    exponents, corresponding to the {isolated} modes.
  }
  \label{tab:floquet_ppo1}
  \begin{tabular}{l l c | l l c}
    \multicolumn{3}{c |}{$\cycle{pp}_{10.25}$} & \multicolumn{3}{c}{$\cycle{rp}_{16.31}$}\\
    $i$ & ~~~~~$\eigRe[i]$  & $\theta_{i}$  & $i$ & ~~~~~$\eigRe[i]$ & $\theta_{i}$  \\
    \hline
    1,2 & ~0.033209  &    $\pm$2.0079  &  1 &     ~0.32791  &              \\
    3 & -4.1096e-13  &                 &  2 &   ~2.8679e-12  &              \\
    4 & -3.3524e-14  &    -1           &  3 &   ~2.3559e-13  &              \\
    5 &  -0.21637    &                 &  4 &     -0.13214  &        -1    \\
    6,7 &  -0.26524  &   $\pm$2.6205   &  5,6 &   -0.28597  & $\pm$2.7724  \\
    8 &  -0.33073    &    -1           &  7 &     -0.32821  &       -1     \\
    9 &  -1.9605    &                  &  8 &      -0.36241  &             \\
    10 & -1.9676    &    -1            &  9,10 &   -1.9617  &  $\pm$2.2411 \\
    $\cdots$ &  $\cdots$    & $\cdots$ & $\cdots$ & $\cdots$ & $\cdots$   \\
    59 &  -5313.6   &    -1           &  59 &   -5314.4 &                 \\
    60 &  -5317.6   &                 &  60 &   -5317.7 &                 \\
    61 &  -6051.8   &    -1           &  61 &   -6059.2 &                 \\
    62 &  -6080.4   &                 &  62 &   -6072.9 &                 \\
    \hline
\end{tabular}
\end{table}

\KSe\ is equivariant under reflection and space translation: $-u(-x,t)$ and
$u(x+l,t)$ are also solutions if $u(x,t)$ is a solution, which corresponds
to equivariance of \eqref{eq:fourierspace} under group operation
 $R=\diag(-1,1,-1,1,\cdots)$ and $g(l)=\diag(r_{1},r_{2},\cdots,r_{N/2-1})$,
where
\[
r_{k}=
\begin{pmatrix}
  \cos(q_{k}l) & -\sin(q_{k}l) \\
  \sin(q_{k}l) & \cos(q_{k}l)
\end{pmatrix}
,\quad k=1,2,\cdots,N/2-1
\,.
\]
Based on the consideration of these symmetries,
there are three types of invariant orbits in \KS\ system: \po s in the
$b_k=0$ invariant antisymmetric subspace, pre\po s which are self-dual
under reflection, and \rpo s with a shift along group orbit after one
period. As shown in \refref{SCD07}, the first type is absent for domains
as small as $L=22$, and thus we focus on the last two types of orbits.
For pre\po s $\hat{u}(0)=R\hat{u}(\period{p})$ , we only need to evolve
the system for a prime period $\period{p}$ which is half of the whole
period, with the Floquet matrix given by
$\jMps_{p}(\hat{u})=R\jMps^{\period{p}}(\hat{u})$. A \rpo,
$\hat{u}(0)=g_p\hat{u}(\period{p})$, returns after one period
$\period{p}$ to the initial state upon the group transform
$g_p=g(l_p)$, so the corresponding Floquet matrix is
$\jMps_p(\hat{u})=g_p\jMps^{\period{p}}(\hat{u})$. Here we show how {\ped} works
by applying it to one representative pre\po\ $\cycle{pp}_{10.25}$
and two
\rpo s $\cycle{rp}_{16.31}$ and $\cycle{rp}_{57.60}$ 
(subscript indicates the period of the orbit), described in
\refref{SCD07}.

\refFig{fig:ppo1rpo1} shows the time evolution of $\cycle{pp}_{10.25}$ 
and $\cycle{rp}_{16.31}$ and the Floquet spectrum of $\cycle{pp}_{10.25}$.
At each repeat of the prime period, $\cycle{pp}_{10.25}$ is invariant
under reflection along $x=L/2$, \reffig{fig:ppo1rpo1}\,(a), and
$\cycle{rp}_{16.31}$ has a shift along the $x$ direction as time goes on,
\reffig{fig:ppo1rpo1}\,(b). Since $\cycle{pp}_{10.25}$ and
$\cycle{rp}_{16.31}$ are both
time invariant and equivariant under SO(2) group transformation $g(l)$,
there should be two marginal Floquet exponents, corresponding to the
velocity field $v(x)$ and group tangent $t(x)=\mathbf{T}x$ respectively,
where $\mathbf{T}$ is the generator of SO(2) rotation:
\[
\mathbf{T}=\diag(t_{1},t_{2},\cdots,t_{N/2-1}),\quad
t_{k}=
\begin{pmatrix}
  0 & -q_{k} \\
  q_{k} & 0
\end{pmatrix}
\,.
\]
\refTab{tab:floquet_ppo1} shows that the $2_{nd}$ and $3_{rd}$,
respectively $3_{rd}$ and $4_{th}$ exponents of $\cycle{rp}_{16.31}$,
respectively $\cycle{pp}_{10.25}$, are marginal, with accuracy as low as
$10^{-12}$, to which the inaccuracy introduced by the error in the closure of
the orbit itself also contributes. \refTab{tab:floquet_ppo1} and
\reffig{fig:ppo1rpo1}\,(c) show that \psd\ is capable of resolving
Floquet multipliers differing by thousands of orders:
when $N=64$, the smallest Floquet multiplier  for $\cycle{pp}_{10.25}$ is
$|\ExpaEig_{62}| \simeq e^{-6080.4\times 10.25}$. We should know that 
this cannot be achieved if we try to get a single
Jacobian for the whole orbit. 
\refFig{fig:ppo1rpo1}(c) and \reftab{tab:floquet_ppo1} also 
show that for $k\ge 9$, Floquet exponents almost lie on the
curve $(q_k^2 - q_k^4 )$. This is the consequence of 
strong dissipation caused by the linear term in \eqref{eq:ksfourier} for
large Fourier mode index. This feature is observed for all
the other \po s we have experimented, and it is a good indicator of 
existence of a finite dimensional inertial manifold. 
Also, Floquet exponents appear in pairs for large indices simply because
the real and complex part of high Fourier modes have similar contracting 
rate from \eqref{eq:ksfourier}.

\begin{figure}[h]
  \centering
  \begin{minipage}{.115\textwidth}
    \centering \small{\texttt{(a)}}
    \includegraphics[width=\textwidth]{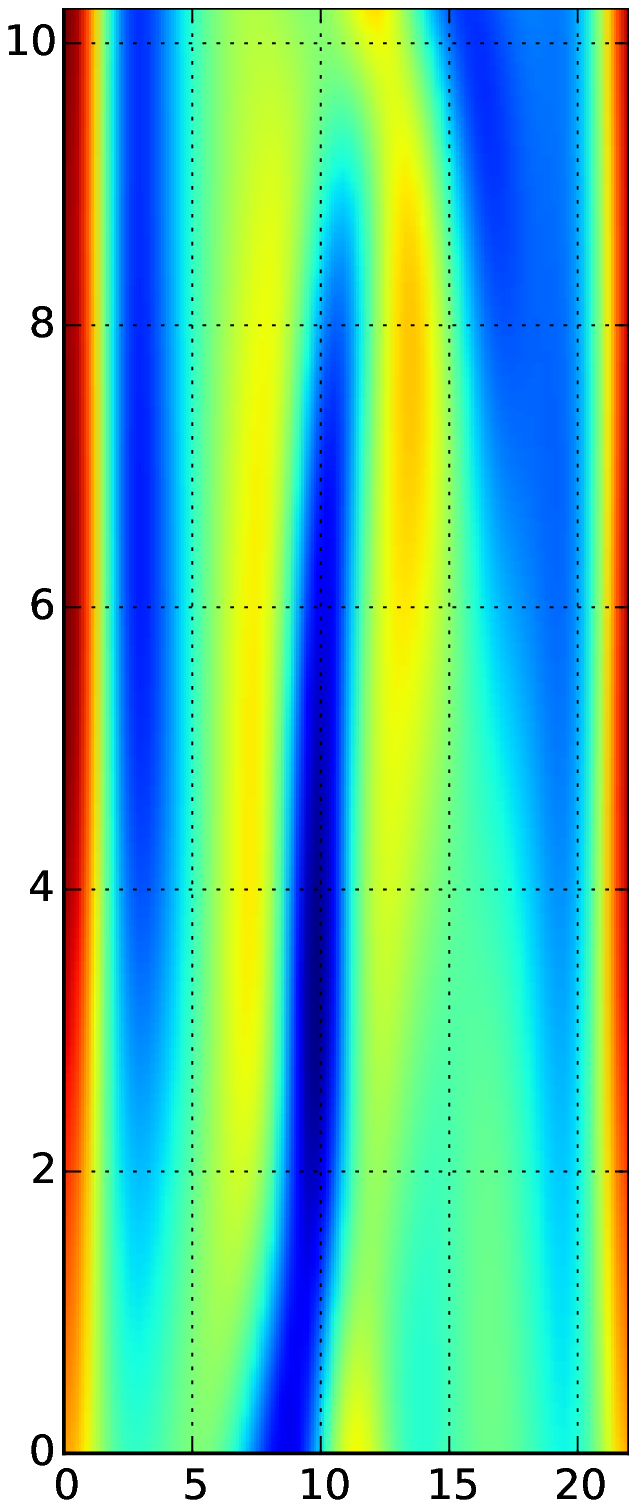}
  \end{minipage}
  \begin{minipage}{.115\textwidth}
    \centering \small{\texttt{(b)}}
    \includegraphics[width=\textwidth]{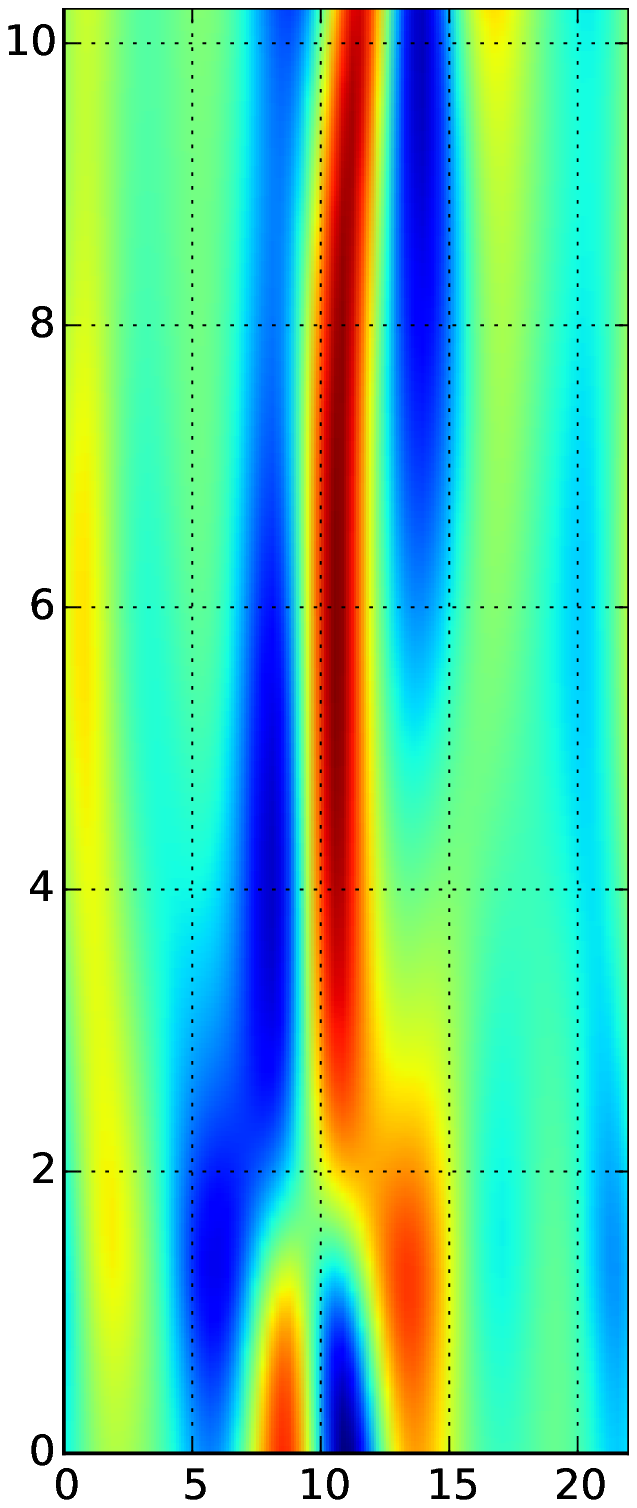}
  \end{minipage}
  \begin{minipage}{.115\textwidth}
    \centering \small{\texttt{(c)}}
    \includegraphics[width=\textwidth]{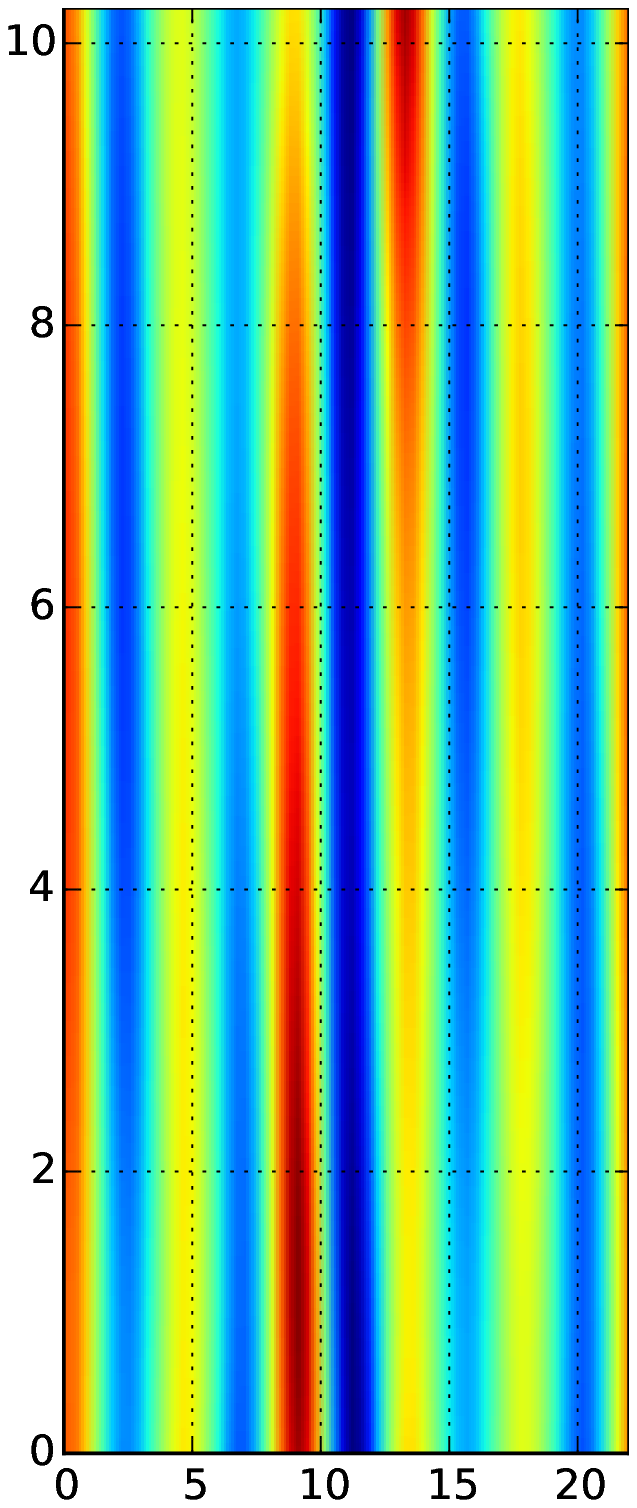}
  \end{minipage}
  \begin{minipage}{.115\textwidth}
    \centering \small{\texttt{(d)}}
    \includegraphics[width=\textwidth]{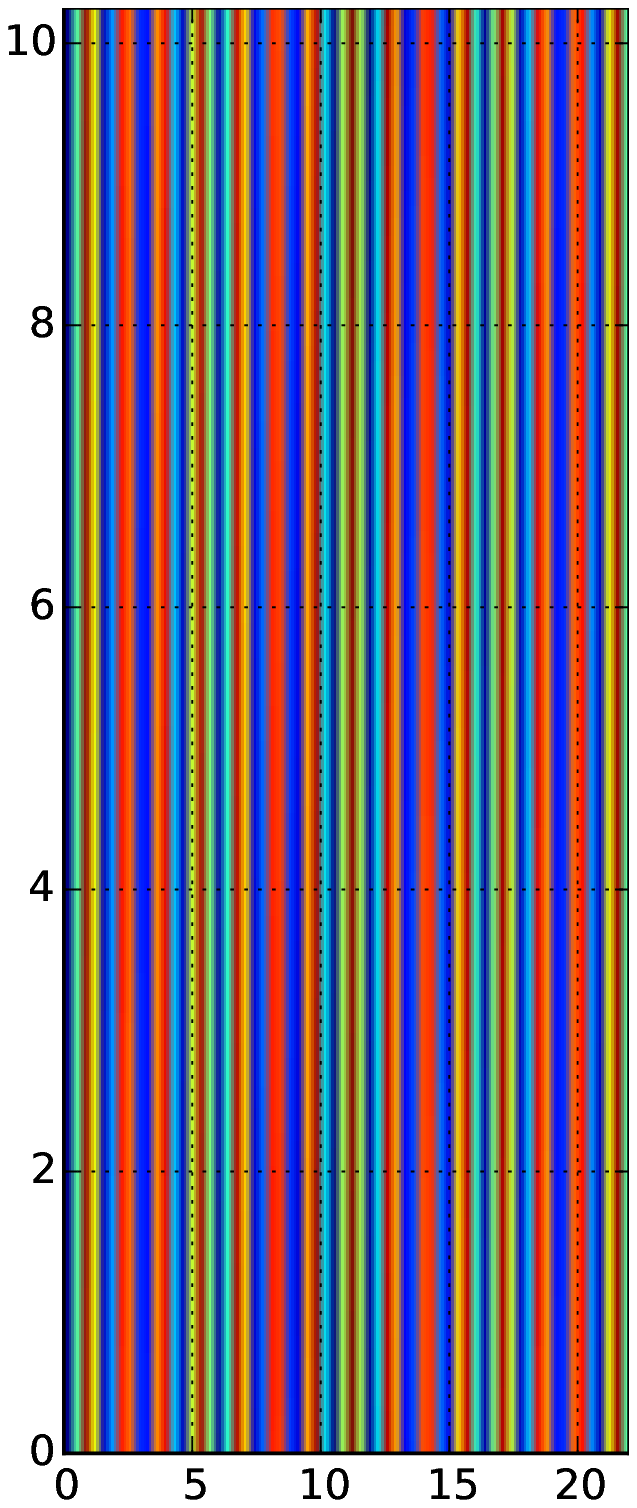}
  \end{minipage}
  \begin{minipage}{.115\textwidth}
    \centering \small{\texttt{(e)}}
    \includegraphics[width=\textwidth]{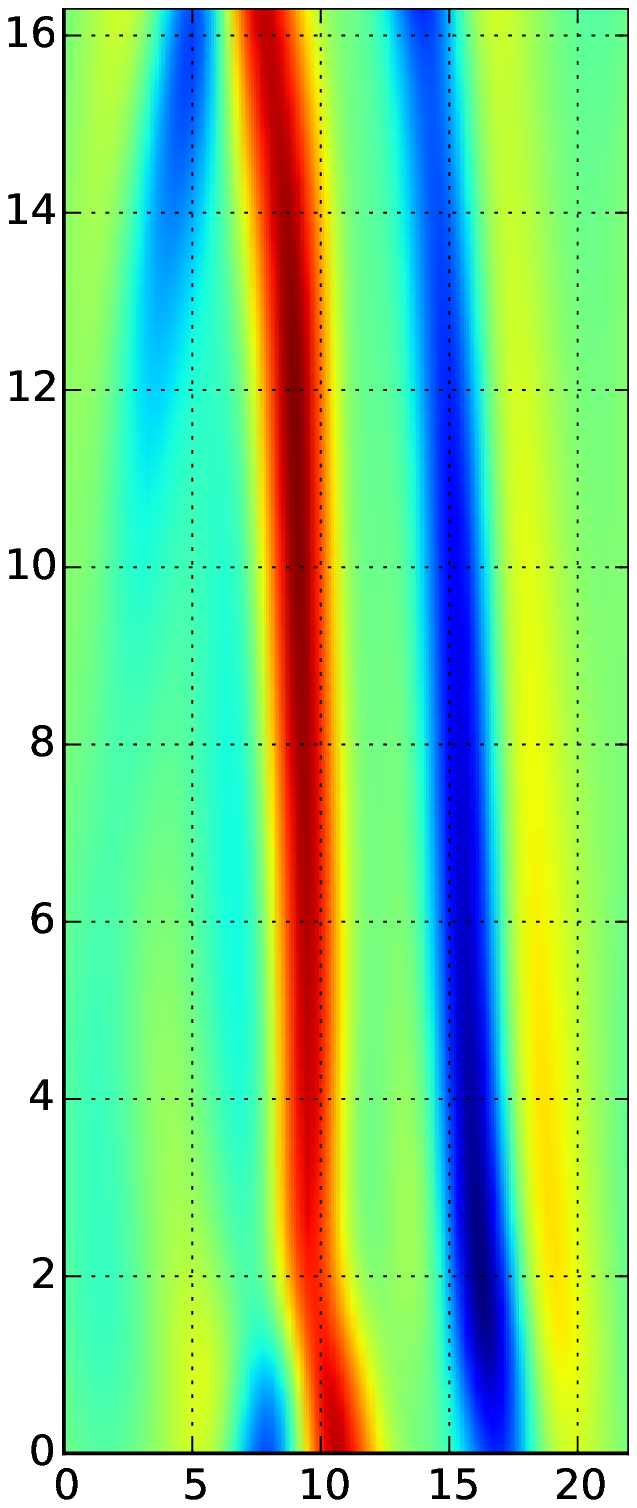}
  \end{minipage}
  \begin{minipage}{.115\textwidth}
    \centering \small{\texttt{(f)}}
    \includegraphics[width=\textwidth]{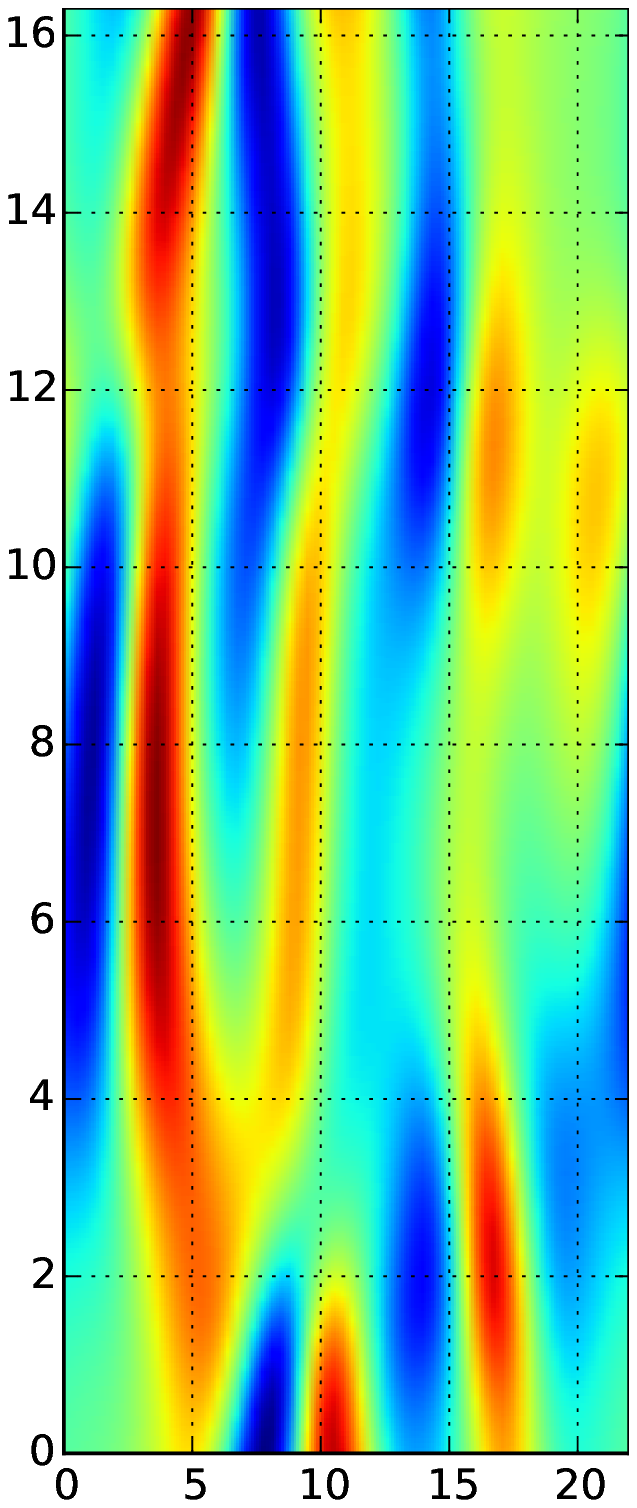}
  \end{minipage}
  \begin{minipage}{.115\textwidth}
    \centering \small{\texttt{(g)}}
    \includegraphics[width=\textwidth]{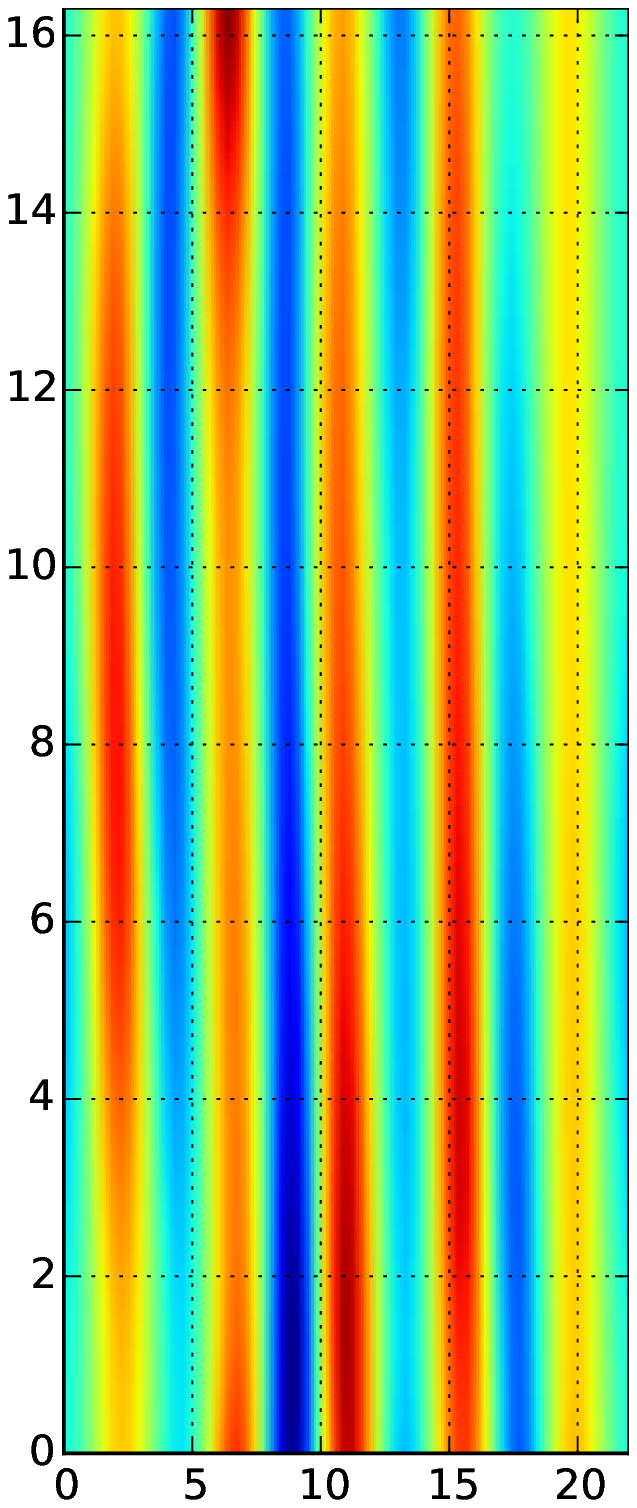}
  \end{minipage}
  \begin{minipage}{.115\textwidth}
    \centering \small{\texttt{(h)}}
    \includegraphics[width=\textwidth]{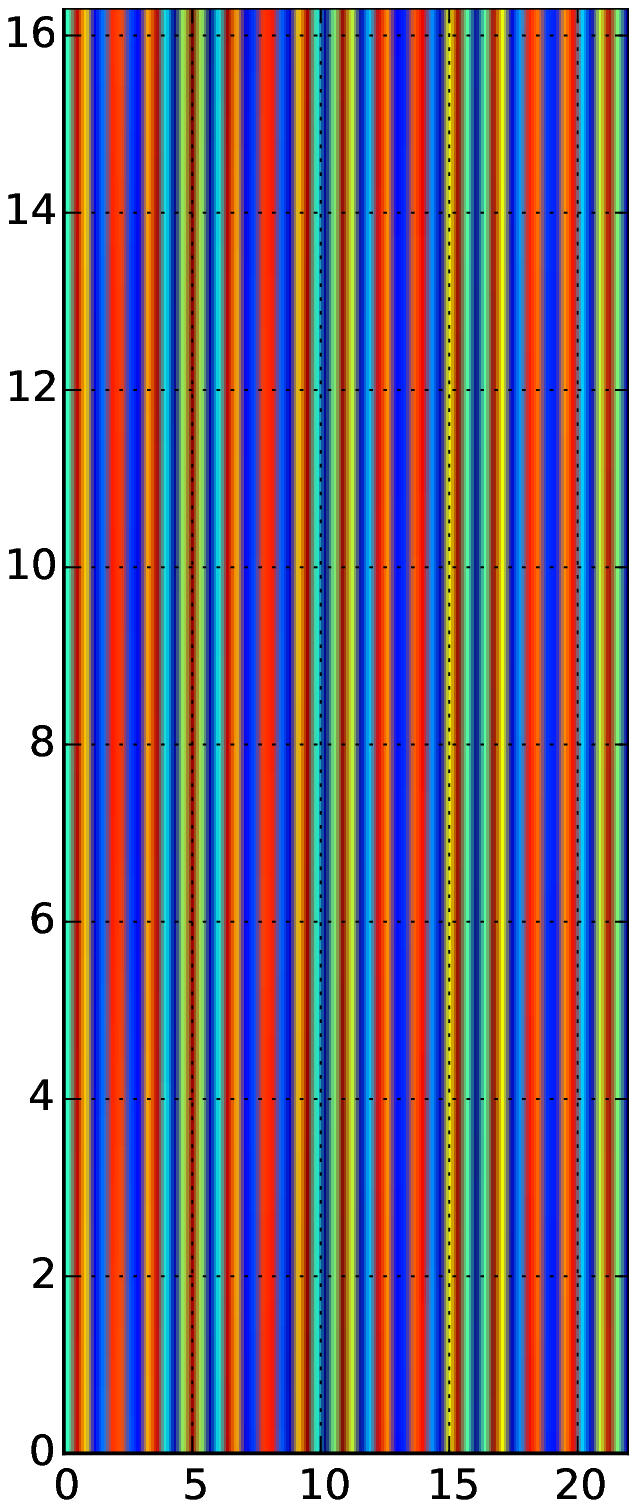}
  \end{minipage}%
  \caption{(Color online)
    (a) $\sim$ (d) : the 1st (real part), 5th, 10th and 30th \Fv\ along 
    $\cycle{pp}_{10.25}$ for one prime period.
    (e) $\sim$ (h) : the 1st, 4th (real part), 10th (imaginary part) 30th (imaginary part) 
    \Fv\ along $\cycle{rp}_{16.31}$ for one prime period.
    axes and color scale are the same with \reffig{fig:ppo1rpo1}.
  }
  \label{fig:Fvs}
\end{figure}
\begin{figure}[h]
  \centering
  \begin{minipage}{.47\textwidth}
    \centering \small{\texttt{(a)}}
    \includegraphics[width=\textwidth]{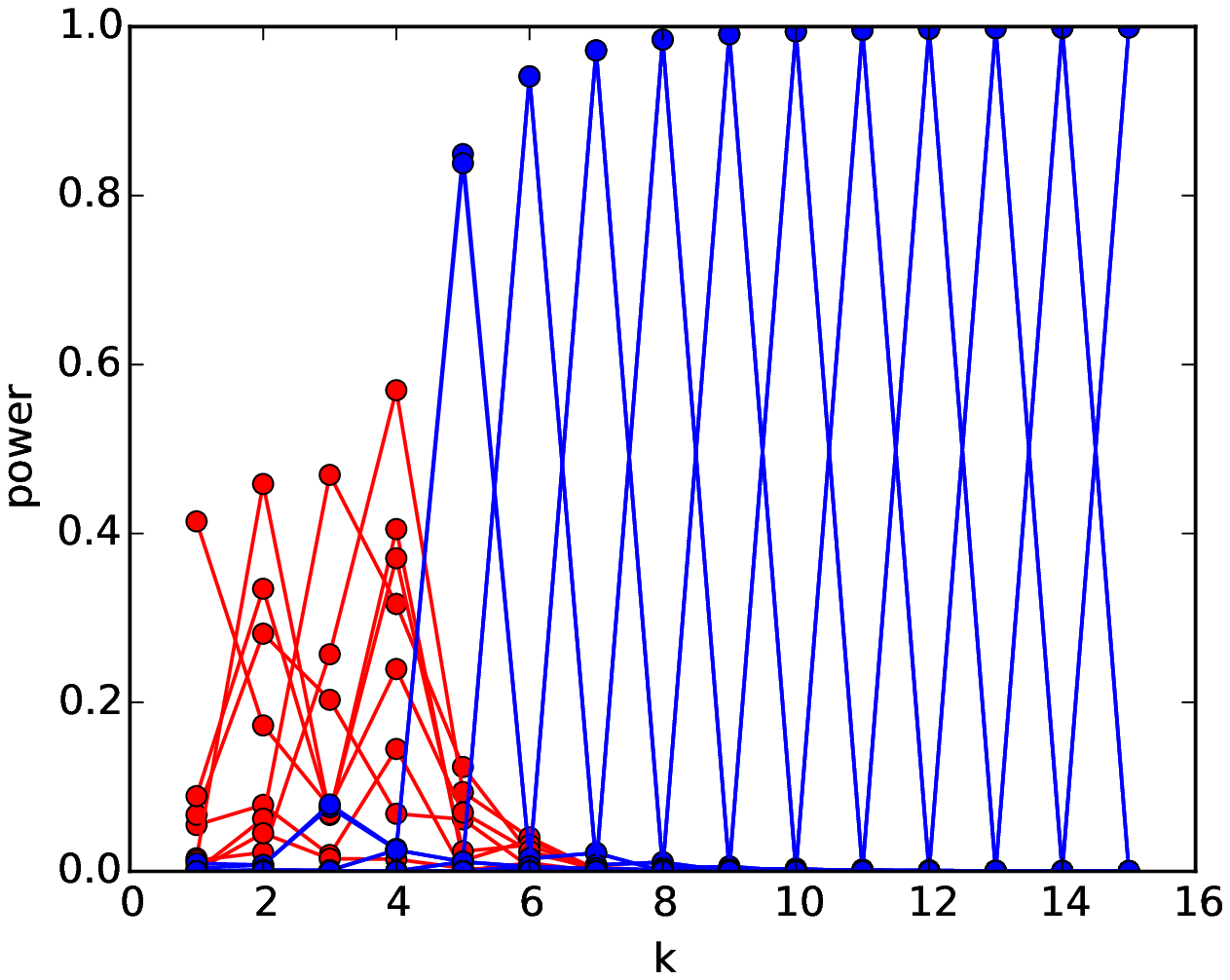}
  \end{minipage}
  \begin{minipage}{.47\textwidth}
    \centering \small{\texttt{(b)}}
    \includegraphics[width=\textwidth]{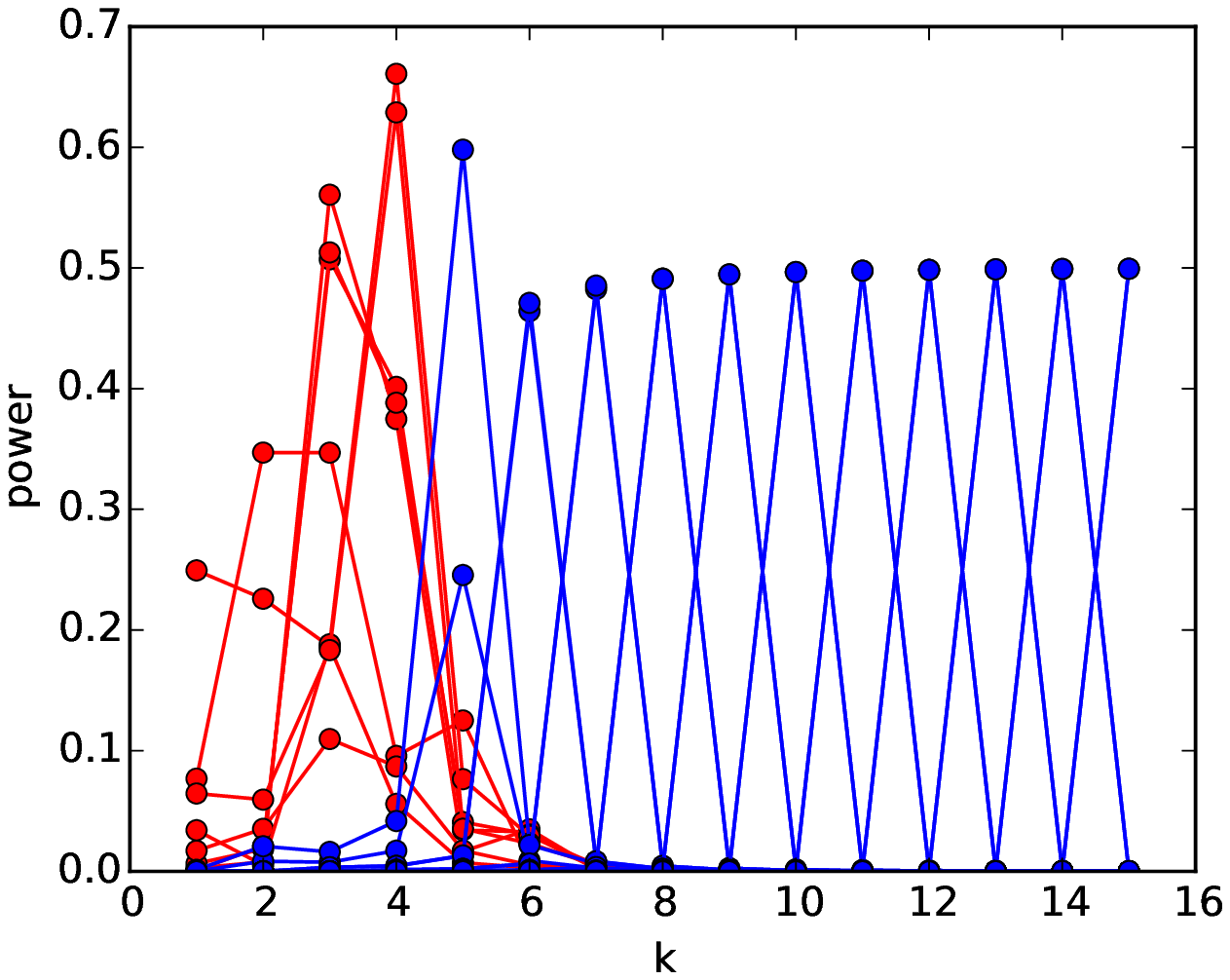}
  \end{minipage}
  \caption{(Color online)
    The power spectrum of the first 30 \Fv s for $\cycle{pp}_{10.25}$ 
    (left)
    and $\cycle{rp}_{16.31}$ (right)
    at $\zeit=0$. Red lines corresponds to the leading 8 \Fv s; while
    the blue lines correspond to the left 22 \Fv s with the $i$th one 
    localized at index $i$.
    Power is defined to
    be the modular square of Fourier coefficients of \Fv s.
    The $x$-axis is
    labeled by the Fourier mode indices. 
    Only the $k>0$ part is shown, the
    negative $k$ follow by reflection. For complex \Fv s, the 
    power spectra of real part and imaginary part are calculated 
    separately. Since almost all contracting \Fv s of $\cycle{rp}_{16.31}$ 
    form complex conjugate pairs, their power peaks are far less than 1.
  }
  \label{fig:FVpower}
\end{figure}
\refFig{fig:Fvs} shows a few selected \Fv s along $\cycle{pp}_{10.25}$
and $\cycle{rp}_{16.31}$ for one prime period respectively. We need to
remind the reader that \Fv s for a whole period is obtained by solving
\eqref{eq:xdpsereal} or \eqref{eq:xdpsdcomplex}, not by evolving 
\Fv s at one time spot to the later time spots because the evolution 
procedure is not stable for \Fv s. We can see that the leading 
few \Fv s have turbulent structures containing only long waves
for both $\cycle{pp}_{10.25}$ and $\cycle{rp}_{16.31}$, but for
\Fv s corresponding to strong contracting rates, the configurations
are pure sinusoidal curves. The power spectra in \refFig{fig:FVpower}
demonstrate this point too. The leading 8 \Fv s have large components in
the first 5 Fourier modes and the spectra are entangled with each other; 
while the remaining \Fv s almost concentrate
on a single Fourier mode and are decoupled from each other;
more specifically, the $i$th \Fv\ with $i\ge 9$
peaks at the $\lceil \frac{i}{2} \rceil$th mode in \reffig{fig:FVpower}.
Takeuchi \etc\rf{TaGiCh11, YaTaGiChRa08} observe similar 
features in \cLv s along ergodic 
trajectories and by measuring the tangency between these two groups of 
\cLv s, they reach a reasonable conclusion about the dimension of 
inertial manifold of \KSe\ and \cGLe.
Therefore, we anticipate that by analyzing the tangency of \Fv s along
different \po s can also lead to the same conclusion, which is our 
further research. 

\begin{figure}[h]
  \centering
  \includegraphics[width=0.47\linewidth]{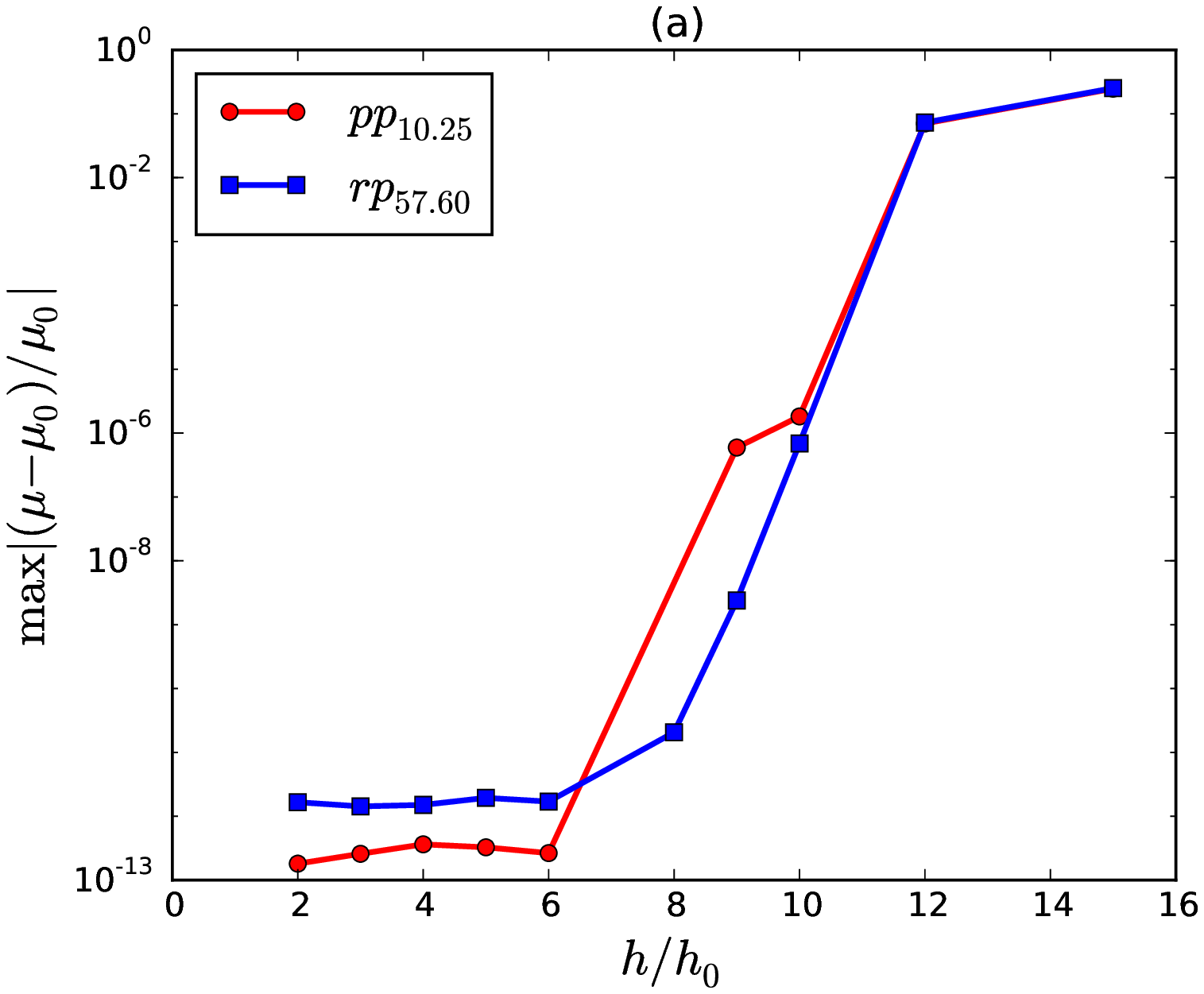} \hfill
  \includegraphics[width=0.47\linewidth]{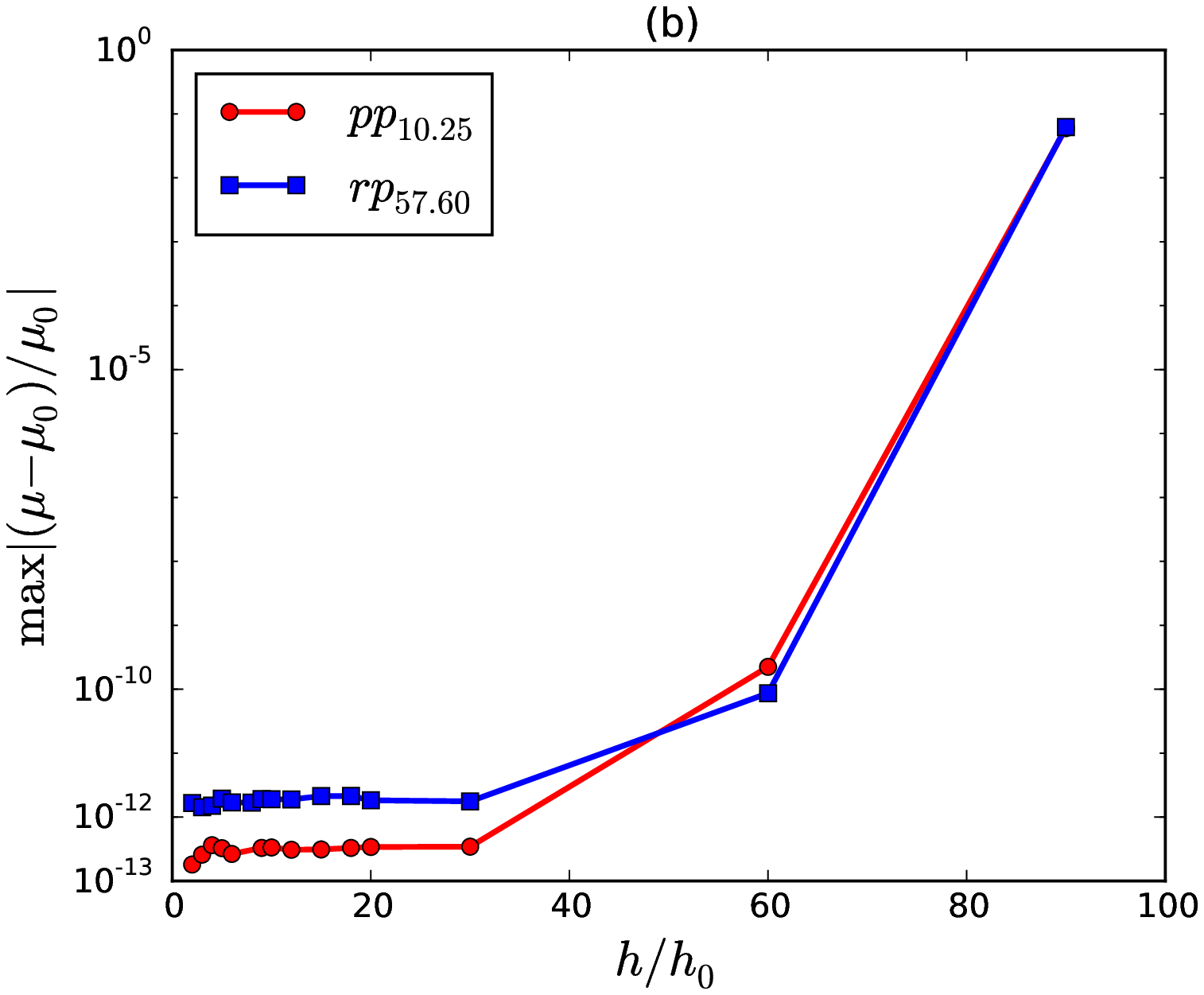}
  \caption{(Color online) Relative error of the real part of
    Floquet exponents associated with different time steps
    with which the Floquet matrix is integrated. Two orbits $\cycle{pp}_{10.25}$
    and $\cycle{rp}_{57.60}$ are used as an example with the base
    case $h_0 \approx 0.001$. (a) The maximal relative difference of
    the whole set of Floquet exponents with increasing time step (decreasing
    the number of ingredient segments of the orbit). (b) Only consider
    the first 35 Floquet exponents.}
  \label{fig:FEerror}
\end{figure}
\begin{figure}[h]
  \centering
  \includegraphics[width=1.0\linewidth]{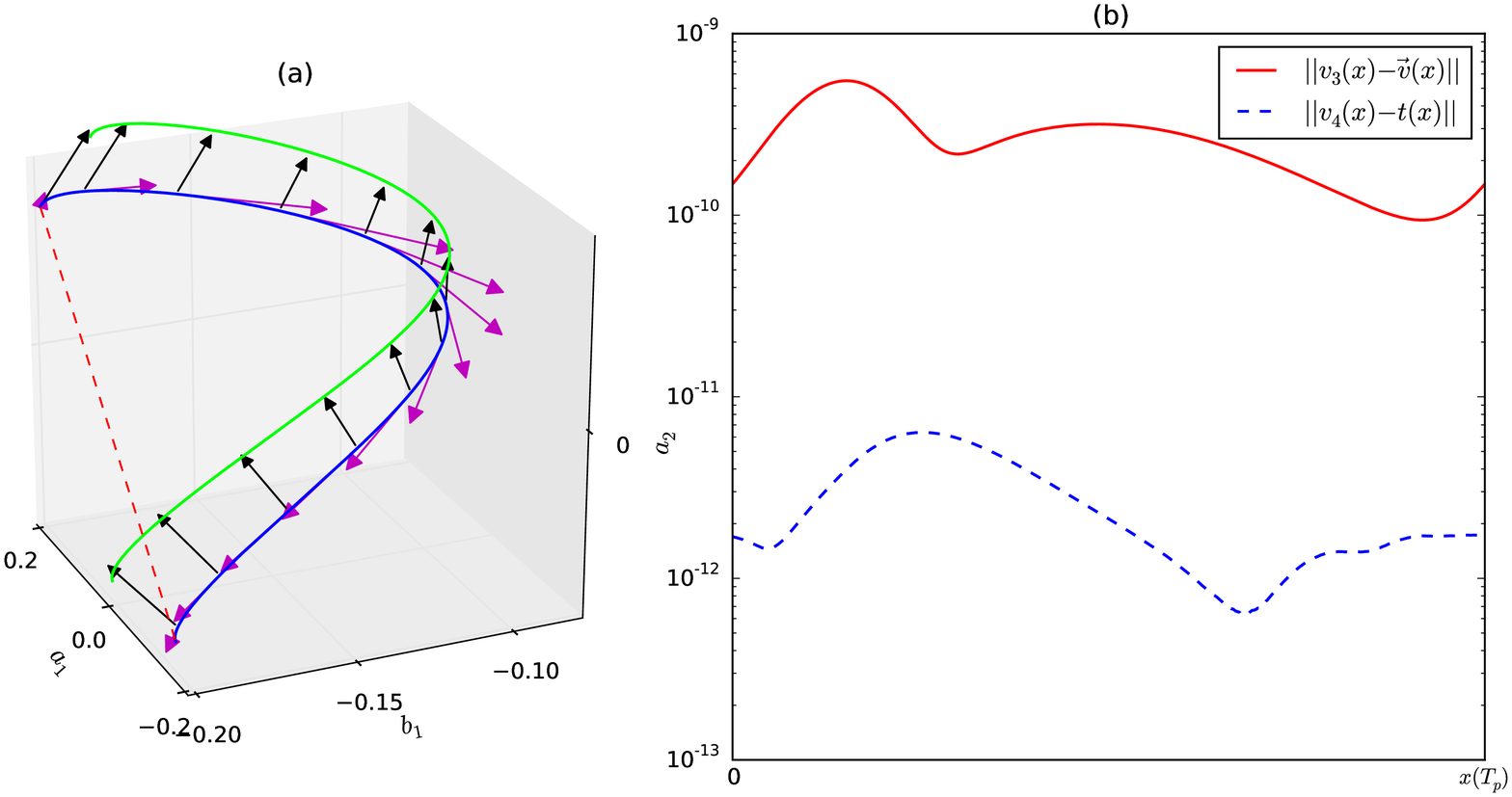}
  \caption{(Color online)
    Marginal vectors and the associated errors.
    (a) $\cycle{pp}_{10.25}$ in one period projected onto $[a_{1},b_{1},a_2]$
    subspace (blue curve), and its counterpart (green line) generated by
    a small group transformation $g(\ell)$
    , here arbitrarily set to $\ell= \,L/(20\pi)$. Magenta and black
    arrows represent the first and the second marginal Floquet vectors
    $\jEigvec[3](x)$ and $\jEigvec[4](x)$ along the prime orbit.
    (b) The solid red curve is the magnitude of the difference between
    $\jEigvec[3](x)$ and the velocity field $\vec{v}(x)$ along the orbit,
    and blue dashed curve is the difference between $\jEigvec[4](x)$ and
    the group tangent $t(x)=\mathbf{T}x$.
  }
  \label{fig:ppo1vectorfield}
\end{figure}
\begin{figure}[h]
  \centering
  \includegraphics[width=0.7\linewidth]{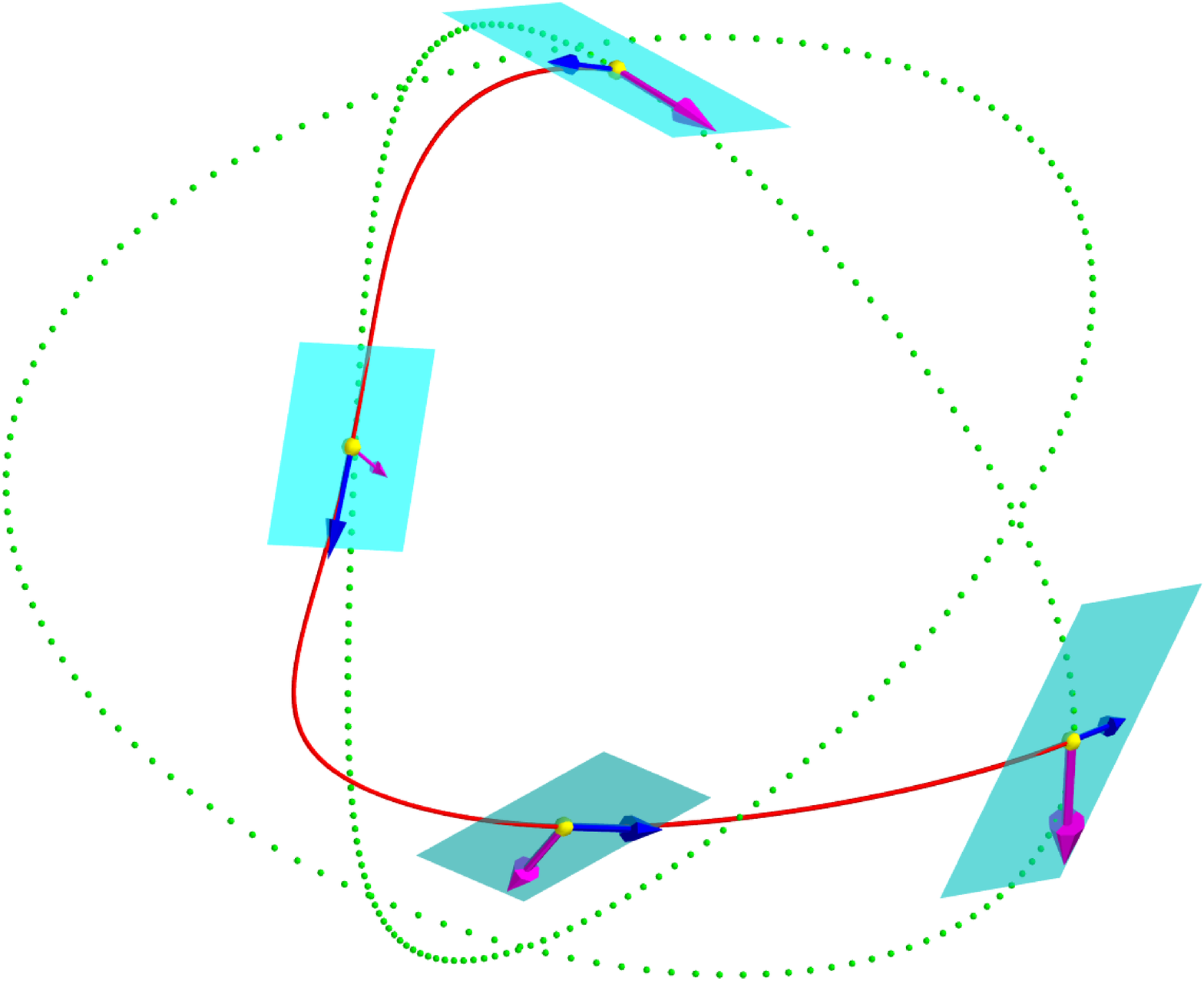}
  \caption{
    (Color online) Projection of \rpo\ $\cycle{rp}_{16.31}$ onto
    the Fourier modes
    subspace $[b_2,c_2,b_3]$ (red curve). The dotted
     curve (lime) is the group orbit
    connecting the initial and final points. Blue and magenta arrows
    represent the velocity field and group tangent along the orbit,
    respectively. Two-dimensional planes (cyan) are spanned by the
    two marginal Floquet vectors at each point (yellow) along the orbit.
  }
  \label{fig:rpo1_marginal3}
\end{figure}

We have noted above that the group property of \JacobianM\
multiplication \eqref{eq:xjacobian} enables us to factorize
$\ps{\jMps}{k}$ into a product of short-time matrices with matrix
elements of comparable magnitudes. In practice, caution should be
exercised when trying to determine the optimal number of time increments
that the orbit should be divided into. If the number of time increments
$m$ is too large, then, according to the estimates of
\refsect{sect:error}, the computation may be too costly. If $m$ is too
small, then the elements of \JacobianM\ corresponding to the
corresponding time increment may range over too many orders of magnitude,
causing \ped\ to fail to resolve the most contracting Floquet vector
along the orbit. One
might also vary the time step according to the velocity at a give point
on the orbit. Here we determined satisfactory $m$'s by numerical
experimentation shown in \reffig{fig:FEerror}. Since larger time step means
fewer time increments of the orbit, a very small time step ($h_0 \approx 0.001$)
is chosen as the base case, and it is increased to test whether the
corresponding Floquet exponents change substantially or not. As shown in
\reffig{fig:FEerror} (a), up to $6h_0$ the whole Floquet spectrum varies within
$10^{-12}$ for both $\cycle{pp}_{10.25}$ and $\cycle{rp}_{57.60}$. These
two orbits represent two different types of invariant solutions which have
short and long periods, so we presume that time step $6h_0$ is good enough
for other short or long orbits too. On the other hand, if only the first
few Floquet exponents are desired, the time step can be increased further
to fulfill the job. As shown in \reffig{fig:FEerror} (b), if we are only
interested in the first 35 Floquet exponents, then time step $30h_0$ is small
enough. In high dimensional nonlinear systems, often we are not interested in the dynamics in the very contracting
directions because they are usually decoupled from the physical modes, and shed little
insight into the system properties. Therefore, large time step could to used in order to
save time.

The two marginal directions have a simple geometrical interpretation and provides
a metric for us to measure the convergence of \ped.
\refFig{fig:ppo1vectorfield}\,(a) depicts the two marginal vectors of
$\cycle{pp}_{10.25}$ projected onto the subspace spanned by $[a_1, b_{1}, a_{2}]$
(the real, imaginary parts of the first mode and the real part of the
second Fourier mode). The first marginal eigen-direction (the $3_{rd}$
Floquet vector in  \reftab{tab:floquet_ppo1}) is aligned with the velocity
field along the orbit, and the second marginal direction (the $4_{th}$
Floquet vector) is aligned with the group tangent. The numerical
difference between the unit vectors along these two marginal directions
and the corresponding physical directions is shown in
\reffig{fig:ppo1vectorfield}\,(b). The difference is under $10^{-9}$ and
$10^{-11}$ for these two directions, which demonstrates the accuracy of
the algorithm.
As shown in \reftab{tab:floquet_ppo1}, for an pre\po, such as $\cycle{pp}_{10.25}$,
the trajectory tangent and the group tangent have eigenvalue $+1$ and
$-1$ respectively, and are thus distinct. However, the two marginal
directions are degenerate for an \rpo, such as $\cycle{rp}_{16.31}$. So these two
directions are not fixed, but the plane that they span is uniquely
determined. \refFig{fig:rpo1_marginal3} shows the velocity field and
group tangent along orbit $\cycle{rp}_{16.31}$ indeed lie in the subspace spanned
by these two marginal directions.

\section{Conclusion and future work}
\label{sect:concl}

In this paper, as well as in the forthcoming publication, \refref{DCTSCD14},
we use one-dimensional \KS\ system to illustrate the effectiveness
and potential wide usage of \ped\ applied to stability analysis
in dissipative nonlinear systems.

On the longer time scale, we hope to apply the method to
the study of orbits of much longer
periods, as well as to the study of high-dimensional, numerically exact
time-recurrent unstable solutions of the full Navier-Stokes equations.
Currently up to 30 Floquet vectors for plane Couette invariant
solutions can be computed\rf{GHCW07}, but many more will be needed
and to a
higher accuracy in order to determine the physical dimension of a turbulent
Navier-Stokes flow. We are nowhere there yet; we anticipate the need for
optimizing and parallelizing such algorithms. Also there is opportunity 
to apply \ped\ to Hamiltonian systems too and we need additional tests to 
show its ability to preserve symmetries of Floquet spectrum imposed by 
Hamiltonian systems.

\section*{Acknowledgments}

We are grateful to L.~Dieci for introducting us to complex
\psd, K.A.~Takeuchi for providing detailed documentation of his previous work
on \cLvs, R.L.~Davidchack for his database of periodic
orbits in \KSe, which are the basis of our numerical experiments,
and to
N.B.~Budanur,  E.~Siminos,  M.M.~Farazmand and H.~Chat\'e
for many spirited exchanges,
X.D. was supported by NSF grant DMS-1028133.
P.C. thanks
G.~Robinson,~Jr.\ for support.

\bibliographystyle{siam}
\bibliography{DingCvit14}

\end{document}